\documentclass[%
 aip,
 amsmath,amssymb,
 reprint,%
]{revtex4-1}

\renewcommand*{\citep}[1]{%
  \begingroup
    \romannumeral-`\x % remove space at the beginning of \setcitestyle
    \setcitestyle{numbers}%
    [\citealp{#1}]%
  \endgroup   
}

\usepackage{graphicx}% Include figure files
\usepackage{dcolumn}% Align table columns on decimal point
\usepackage{bm}% bold math
\usepackage[utf8]{inputenc}
\usepackage[T1]{fontenc}
\usepackage{mathptmx}
\usepackage{siunitx}
\usepackage{float}
\usepackage{placeins}
\usepackage[usenames,dvipsnames]{xcolor}
\definecolor{olive}{rgb}{0.5, 0.5, 0.0}
\definecolor{bluegray}{rgb}{0.4, 0.6, 0.8}
\usepackage{todonotes}
\presetkeys{todonotes}{inline}{}

\newcommand{\etal}{\textit{et al.}}

\begin{document}

\preprint{AIP/123-QED}

\title[]{Electron self-injection threshold for the tandem-pulse laser wakefield accelerator}
% Force line breaks with \\
%%%%
\author{Zahra M. Chitgar}
\email[]{z.chitgar@fz-juelich.de}
%\thanks{}
\affiliation{Institute for Advanced
	Simulation, J\"ulich Supercomputing Centre, Forschungszentrum J\"ulich, D-52425 J\"ulich,
	Germany}
\author{Paul Gibbon}
%\email[]{}
%\thanks{}
\affiliation{Institute for Advanced
	Simulation, J\"ulich Supercomputing Centre, Forschungszentrum J\"ulich, D-52425 J\"ulich,
	Germany}
\affiliation{Centre for Mathematical Plasma Astrophysics, Katholieke Universiteit Leuven, 3000 Leuven, Belgium}

\author{J\"urgen B\"oker}
%\email[]{}
%\thanks{}
\affiliation{Institut f\"ur Kernphysik (IKP-4), Forschungszentrum J\"ulich, D-52425 J\"ulich, Germany}

\author{Andreas Lehrach}
%\email[]{}
%\thanks{}
\affiliation{Institut f\"ur Kernphysik (IKP-4), Forschungszentrum J\"ulich, D-52425 J\"ulich, Germany}
\affiliation{Institut f\"ur Experimentalphysik III B und III. Physikalisches Institut, RWTH Aachen, Germany}

\author{Markus B\"uscher}
%\email[]{}
%\thanks{}
\affiliation{Peter Gr\"unberg Institut (PGI-6), Forschungszentrum J\"ulich GmbH,  D-52425 J\"ulich,
	Germany}
\affiliation{Institut f\"ur Laser- und Plasmaphysik
	Heinrich-Heine Universit\"at D\"usseldorf, D-40225 D\"usseldorf,
	Germany}

\date{\today}% It is always \today, today,

\begin{abstract}
A controllable injection scheme is key to producing high quality laser-driven electron beams and X-rays. Self-injection is the most straightforward scheme leading to high current and peak energies, but is susceptible to variations in laser parameters and target characteristics. In this work improved control of electron self-injection in the nonlinear cavity regime using two laser-pulses propagating in tandem is investigated. In particular the advantages of the tandem-pulse scheme in terms of injection threshold, electron energy and beam properties in a regime relevant to betatron radiation are demonstrated. Moreover it is shown that the laser power threshold for electron self-injection can be reduced by up to a factor of two compared to the standard, single-pulse wakefield scheme.
\end{abstract}

\maketitle

\section{Introduction}\label{intro}
Compact laser-plasma electron accelerators have made enormous strides over the past two decades in terms of beam energy, quality and reproducibility. A key factor in this advance has been the separation of acceleration and injection phases, offering more control over the beam dynamics \cite{malka_cern2016}. 
Apart from providing a route to cheaper, more accessible GeV electron beams, laser-based accelerators are also becoming an important source of  bright X-rays with femtosecond to picosecond pulse duration. These are of great importance in different branches of science for resolving atomic structure down to sub-nanometer range and for capturing ultrashort time scale events \cite{lemke2013femtosecond,rousse2001femtosecond,uesaka2000generation}. Until now synchrotrons and free electron lasers have provided such sources, but these have limited accessibility because of their costs and huge scale. Laser-driven betatron radiation is a relatively recent source of femtosecond X-rays which could potentially offer much wider availability~\cite{rousse2004production,kneip2008observation,kneip2010bright,corde2011controlled}.

The most widely studied scheme is the laser wakefield accelerator (LWFA)\cite{tajima1979laser}, in which plasma electrons are pushed away by the ponderomotive force of a focused laser pulse to regions with lower intensity. Ions remain immobile for such short interaction times, so that the electrons start to oscillate around their initial position due to the restoring force. This perturbation, or wake, is strongest for pulse durations matched to the electron plasma period, or $\tau_L\sim\omega_p^{-1}$, and follows the laser pulse with a phase velocity equal to the group velocity of the laser pulse. In this way, injected electrons surfing on this wave can get accelerated.

Since the LWFA was first conceived, several techniques have been proposed to increase the energy and flux of the injected electrons and their corresponding emitted radiation. Multi-pulse (MP-LWFA) schemes have been advocated before to successively increase the amplitude of the plasma wave and achieve higher electron energies~\cite{umstadter1994nonlinear,dalla1994large,kim2007double} with greater efficiency, and have also been studied in the context of betatron radiation in the quasi-linear regime~\cite{hooker2014multi}. Tailored density profiles \cite{bonnaud1994wake,tomassini2017resonant} and two-color driven ionization injection~\cite{bourgeois2013two,yu2014two,xu2014low} have been explored as a means of reducing the energy spread and emittance of the electron beam. Recently, a combined cluster/gas-jet target was shown to yield higher betatron radiation flux and energy~\cite{chen2013bright} than a standard gas target.

In this paper we show how the injection process in the nonlinear bubble regime can be better controlled using two co-propagating pulses with the same wavelength and focal lengths but differing intensity. This scheme results in a lower injection threshold in terms of the laser intensity, while at the same time yielding improved beam qualities such as energy, charge and emittance compared to the single-pulse scheme, potentially offering advantages for TW laser-driven betatron radiation sources. These findings may also make it easier to extend the recent progress in generation of compact MeV electron beams to a new class of kHz laser facilities now coming into use \cite{guenot_natphot2017,JUSPARC_fzj2019}.

In order to focus the investigation on the role of the second pulse in the injection process, we modelled a homogeneous gas target where the lasers are focused far from the edges and a simulation time limited to around one Rayleigh length. This choice is designed to reduce the influence of other injection-triggers such as density gradients and nonlinear pulse propagation effects such as self-focusing, which will likely come into play at later times.

The paper is organized as follows. In Sec.~\ref{injection}, the fundamentals of electron injection and different means of achieving it are briefly introduced, including the nonlinear cavity regime of electron acceleration. In Sec.~\ref{result} the simulation methodology is described and the optimal conditions for the double-pulse scheme with respect to the electron beam properties are evaluated. In Sec.~\ref{injthreshold} the injection threshold using the double pulse scheme is determined and compared to a single-pulse driver for experimentally relevant conditions. The paper concludes with a discussion on the merits and practicalities of this scheme.

\section{Electron self-injection in laser-driven accelerators}\label{injection}

The `blowout' or `bubble'  regime was first predicted for electron-beam-driven~\cite{rosenzweig_pra91} and laser-driven~\cite{pukhov2002laser} plasma wakefields respectively, and later analyzed~\cite{kostyukov2004,lu2006nonlinear} as the highly non-linear regime of electron wakefield acceleration. Lu et al.\cite{lu2007generating} demonstrated that `matched' cavity formation can be achieved by balancing the transverse ponderomotive force of the laser to the radial space-charge force created by the ion channel. This yields a condition for which the focal spot-size $w_0$ is comparable to the blowout radius, ie: $k_pw_0 \simeq k_pr_b \simeq 2\sqrt{a_0}$, for laser amplitudes $a_0\ge 2$. The pulse duration should also be roughly a plasma period, $\tau_L=\omega_p^{-1}$, where $\omega_p=ck_p$ is the cold plasma frequency. Here $\lambda_p=2\pi c/k_p$ denotes the plasma wavelength, $a_0 = \SI{8.5E-10}{}(I_L[W/cm^2])^{1/2}\lambda_L[\mu m]$ the dimensionless laser field strength for laser intensity $I_L$, and $\tau_L$ the pulse length.

Under certain conditions, some plasma electrons can get trapped at the back of this bubble and be accelerated in the strong longitudinal electric field of the ion cavity~\cite{esirkepov2006electron}. Moreover, its radial electric field causes electrons start to oscillate around the laser propagation axis. This motion results in betatron radiation; emission of bright X-rays directed according to the energy and trajectory of the oscillating electrons~\cite{corde2013femtosecond}. This implies that the radiation flux can be tuned through the electron dynamics, for which the injection process and the cavity structure both play crucial roles.

In the simplest one-dimensional picture, the minimum energy of the electrons to get injected into a laser-driven wakefield is required to be greater than the wake phase velocity, which to a first approximation is tied to the group velocity of the driver pulse\cite{esarey_pop95}, $p_{min}/m_ec\approx\gamma_{p}\simeq \omega_0/\omega_p$. This condition can be met in different ways: either through modification of the (generally three-dimensional) wakefield structure so that some of the plasma wake electrons get injected, or by injecting additional background electrons into a pre-formed wake. Wave-breaking and formation of cavity provide the necessary conditions for betatron oscillations, but electron injection does not occur automatically. To inject the electrons at the right point in the wakefield, several techniques have been considered and implemented so far, which all have their advantages and drawbacks~\cite{CERNYR218}. 

In order to obtain high quality electron beams for applications, a low-emittance, narrow-energy-spread electron bunch is required, meaning that control over the electron injection process is essential. An attractive aspect of the cavity regime is that self-injection readily occurs provided that -- in addition to the matching requirements introduced earlier -- the laser power $P$ satisfies $P/P_c \gtrsim 5$, where $P_c = 17.4(\omega_0/\omega_p)^2$~GW is the critical power for self-focusing~\cite{tsung2006simulation,lu2007generating}.  This rule of thumb has been largely confirmed in experiments, but these also show that the electron beam quality thus obtained is very erratic\cite{CERNYR218,froula2009measurements,mangles2012self}. To improve reproducibility of beam properties therefore, various injection aids have been considered such as: ionization injection~\cite{pak2010injection,mcguffey2010ionization,clayton2010self}, optical injection~\cite{faure2006controlled,rechatin2009controlling,lundh2011few}, or modification of the plasma wave dispersion properties via density gradients~\cite{brantov2008controlled,geddes2008plasma,schmid2010density,he2013high,buck2013shock}. 

In this paper we investigate the threshold conditions and bunch quality for an all-optical, double-pulse injection scheme, in which the nonlinear cavity field is amplified in a controlled manner via a 2nd trailing pulse, and the injected electron bunch current is enhanced by charge accumulation at the rear of the second cavity thanks to recycling of electrons initially expelled by the first pulse. We note in passing that similar collinear pulse configurations have also been considered in the context of ionization injection \cite{bourgeois2013two}, two-color driven wakefield~\cite{pathak2018all,Kalmykov_2018} and channel-guided wakefield acceleration \cite{thomas_prl2008}, where the latter experiment used a wider, lower-intensity leading pulse to create a plasma channel. By contrast, the scheme studied here assumes that the laser pulses share the same focal optics and enter a homogeneous gas target.

\section{Double pulse Scheme}\label{result}
Such a collinear double-pulse scheme was recently examined by Horn\'{y} \etal~\cite{horny_ppcf2018}, using a lower-intensity preceding laser pulse as a means of gaining more control over the self-injected bunch charge and dynamics, potentially leading to higher energy gain and/or bunch charge for electrons and enhanced betatron emission in the nonlinear blowout regime~\cite{horny2017,chitgar2018}. 
This tandem-pulse scheme, which relies on a double-cavity, has two advantages: first, the leading pulse ensures that the plasma encountered by the 2nd driver is fully pre-ionized; second, the accumulation and recycling of free electrons at the back of the first cavity provides a concentrated source for the second pulse to act on, ultimately enhancing the accelerating field behind it, resulting in higher electron beam energy and current. These observations naturally raise the question of when it becomes worthwhile to divide the available laser energy into such an 'injector-driver' configuration, and if so, how their relative intensities and timing need to be arranged. In the following sections we attempt to examine this question before making a more systematic assessment of the scheme compared to the conventional single-pulse drive mode.

\begin{figure*}[ht]
	\includegraphics[scale=0.5]{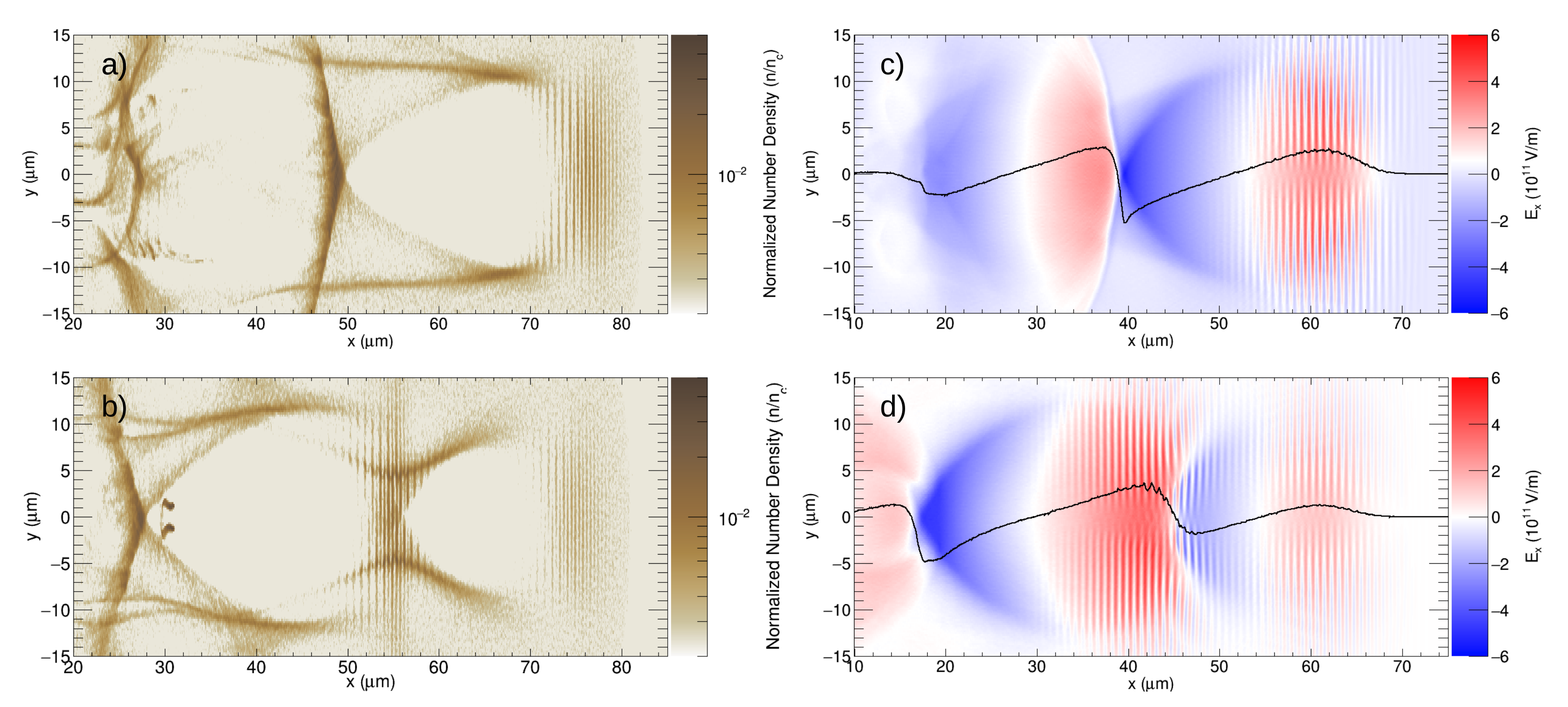}
	\caption{Comparison between single- and double-pulse schemes for optimal conditions (see Sections~\ref{delayopt} and \ref{energyopt}) with $(\gamma_p,a_0)=(30,5)$. Left: 2D snapshot of the electron number density, where the target is irradiated by a laser with 'total' pulse energy of \SI{3.4}{\joule}, for a) the single pulse scheme and b)
		the double-pulse scheme with optimum condition, at $t=\SI{1300}{\femto\second}$. The right-hand side of the figure shows the longitudinal component of electric field just before injection commences at \SI{290}{\femto\second} for c) the single pulse scheme, and d) the double pulse scheme. }
	\label{numdens} 

\end{figure*}

\subsection{Simulation method}

To make a quantitative survey covering a range of realistic parameters we have performed both 2-dimensional and 3-dimensional particle-in-cell simulations using the EPOCH code~\cite{arber2015contemporary}.
All 2D (3D) simulations were performed using a $100\times \SI{80}{\micro\meter\squared}$ $(100\times 80 \times\SI{80}{\cubic\micro\meter})$ box filled with an underdense helium gas (preionized helium plasma), discretised by a computational grid with dimensions $n_x\times n_y = 3125\times 400$ $(n_x\times n_y \times n_z = 3125\times 400\times 400)$ and 2 (4) particles per cell. A number of higher-resolution simulations were also done to check statistical convergence with no significant change in the results. The target is preceded by a vacuum region of \SI{5}{\micro\meter} followed by a \SI{12}{\micro\meter} ramp in the gas density in order to avoid an overly steep gradient at the plasma edge. A $\SI{20}{\femto\second}$ laser pulse with wavelength  $\SI{800}{\nano\metre}$ was focused from the left hand boundary down to a $\SI{10}{\micro\metre}$ $1/e$ focal waist $w_0$ (FWHM$=\SI{16.6}{\micro\metre}$ ) at the box center. A moving window was deployed in order to follow the development of the ensuing plasma wake and electron injection, which is switched on $\SI{260}{\femto\second}$ after the start of simulation, before the laser pulse reaches the right side of the boundary. For the 2D runs, the amount of charge fully injected into the cavity is estimated from the number of particles extrapolated into a 3D box surrounding the injected bunch using a fifth order (B-Spline) particle weighting. 

Most simulations were run up to a time corresponding to the Rayleigh length of the laser pulse ($z_R=\pi w_0^2/\lambda$), allowing a well-defined early-time assessment of the double-pulse scheme compared to its single-pulse equivalent. Limiting the simulation time in this way also minimizes additional propagation effects (refraction, focussing or etching) which might influence the injection process at later times~\cite{kalmykov2009electron}. We will return to this point later in Section~\ref{discussion}.

A representative example of the new scheme is displayed in Fig.~\ref{numdens}, which compares the electron number density after $\textrm{t}=\SI{1300}{\femto\second}$ for the single- and double-pulse schemes respectively, corresponding to approximately one Rayleigh length  of laser propagation; in this example $\sim \SI{400}{\micro\meter}$. The target density is $\num{1.9E18}\SI{}{\per\cubic\centi\metre}(\gamma_p=30)$ which is irradiated by a laser pulse corresponding to $a_0=5$ for both single- and double-pulse schemes, meaning that the same total energy of $\mathrm{U}_{\mathrm{total}}=\SI{3.4}{\joule}$ is used for both cases, but \textit{shared} between pulses in the double pulse scheme ($\mathrm{U}_2/\mathrm{U}_{\mathrm{total}} = 0.84, \mathrm{U}_1/\mathrm{U}_{\mathrm{total}} = 0.16$).

In the double-pulse scheme the second cavity is slightly longer and wider than the first cavity of the single-pulse case in Fig.~\ref{numdens}(a,b). Consequently a longer acceleration length could be expected, as well as a potentially larger cross-section for injected particles. The latter is confirmed by the trajectories shown later in  Fig.~\ref{trajectory}(a), which trace the origin of the electron bunches injected into the second cavity seen in Fig~\ref{numdens}(b). With the same laser parameters and target characteristics, there is no electron injection in the first bubble in the single-pulse scheme --- Fig.~\ref{numdens}(a). This indicates that for the same total laser energy, the injection threshold appears to be markedly reduced for the tandem scheme compared to a single pulse driver.

\begin{figure}[ht]
	\includegraphics[scale=0.67]{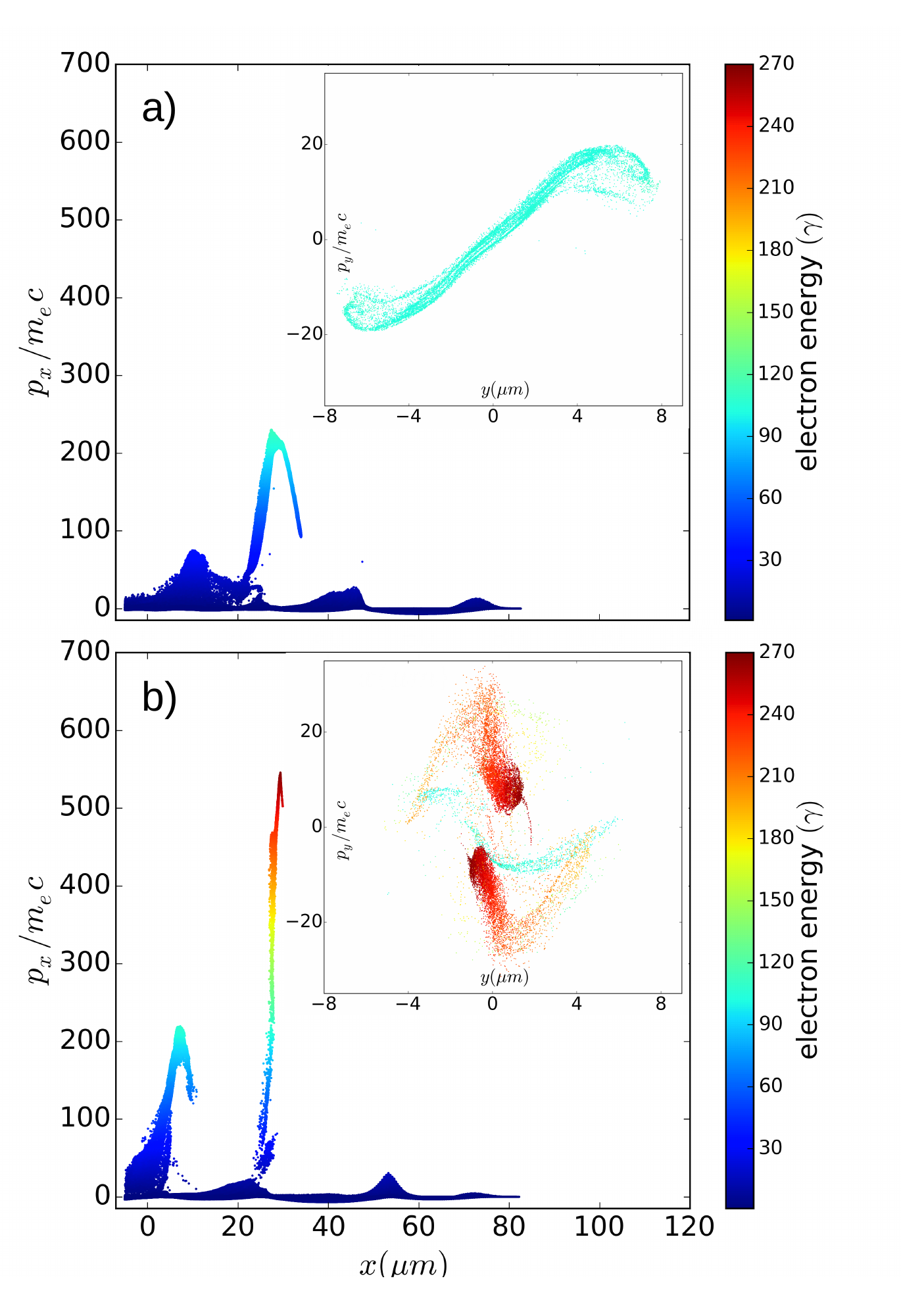}
	\caption{Longitudinal momentum phase space $(x,p_x)$ at $t=\SI{1300}{\femto\second}$ for a) single-pulse and b) double-pulse scheme with optimum condition for $(\gamma_p,a_0) = (30,5)$, corresponding to the simulations in Fig.~\ref{numdens}. The small inset in each figure shows the transverse momentum phase space $(y,p_y)$.  }
	\label{emit} 

\end{figure}

Figure~\ref{emit} depicts the electron distribution in longitudinal $(x,p_x)$ momentum phase space for (a) single- and (b) double-pulse scheme in Fig.~\ref{numdens}, and the small inset in each figure shows the transverse $(y,p_y)$ momentum phase space. In this example the injected electrons in the second cavity of the double-pulse scheme carry around $ \SI{48}{\pico\coulomb}$ with a normalized emittance of $\SI{7.86\pi}{\milli\metre\milli\radian}$, accelerated up to $ \SI{140}{\mega\electronvolt}$. In the single-pulse case, there is no injection in the first cavity. However, there is a bunch of electrons at the back of the first cavity carrying $\SI{228}{\pico\coulomb}$ with a normalized emittance of $\SI{23.28\pi}{\milli\metre\milli\radian}$. This bunch is accelerated up to $ \SI{50}{\mega\electronvolt}$, within the same target length and the pulse energy. Although this latter bunch is carrying a much higher amount of charge, the quality of bunch is significantly degraded by the onset of cavity wall deformation.

A key characteristic of bubble acceleration is that a volume devoid of electrons is created by the first pulse. On the other hand, expelled electrons return back to the rear side of the bubble providing surplus concentrated charge for the second pulse to resonantly act on in creating a second, stronger cavity --- Fig.~\ref{numdens}(a,b). The advantage of double pulse scheme over single pulse is also apparent by the higher electric field strength in (d) compared to (c) extended over a longer distance: $\Delta(E.d)\sim \SI{1.0}{\mega\volt}$ and $\SI{0.7}{\mega\volt} $, respectively. 

An initial set of simulations was performed  aimed at optimizing the double-pulse scheme, before comparing the injection threshold in single- and double-pulse schemes. 
These simulations were done for a specific laser pulse with amplitude $a_0 = 5$ (corresponding to a total energy of \SI{3.4}{\joule}) and helium gas density of $\num{1.9E18}\SI{}{\per\cubic\centi\metre}$, the details of which follow in Sec.~\ref{delayopt} and~\ref{energyopt}. A parameter scan of injection threshold simulations for six laser amplitudes and 5 different target densities are described later in Sec.~\ref{injthreshold}. 

\subsection{Optimization of Pulse delay}\label{delayopt}
\begin{figure}[ht]
	\includegraphics[scale=0.47]{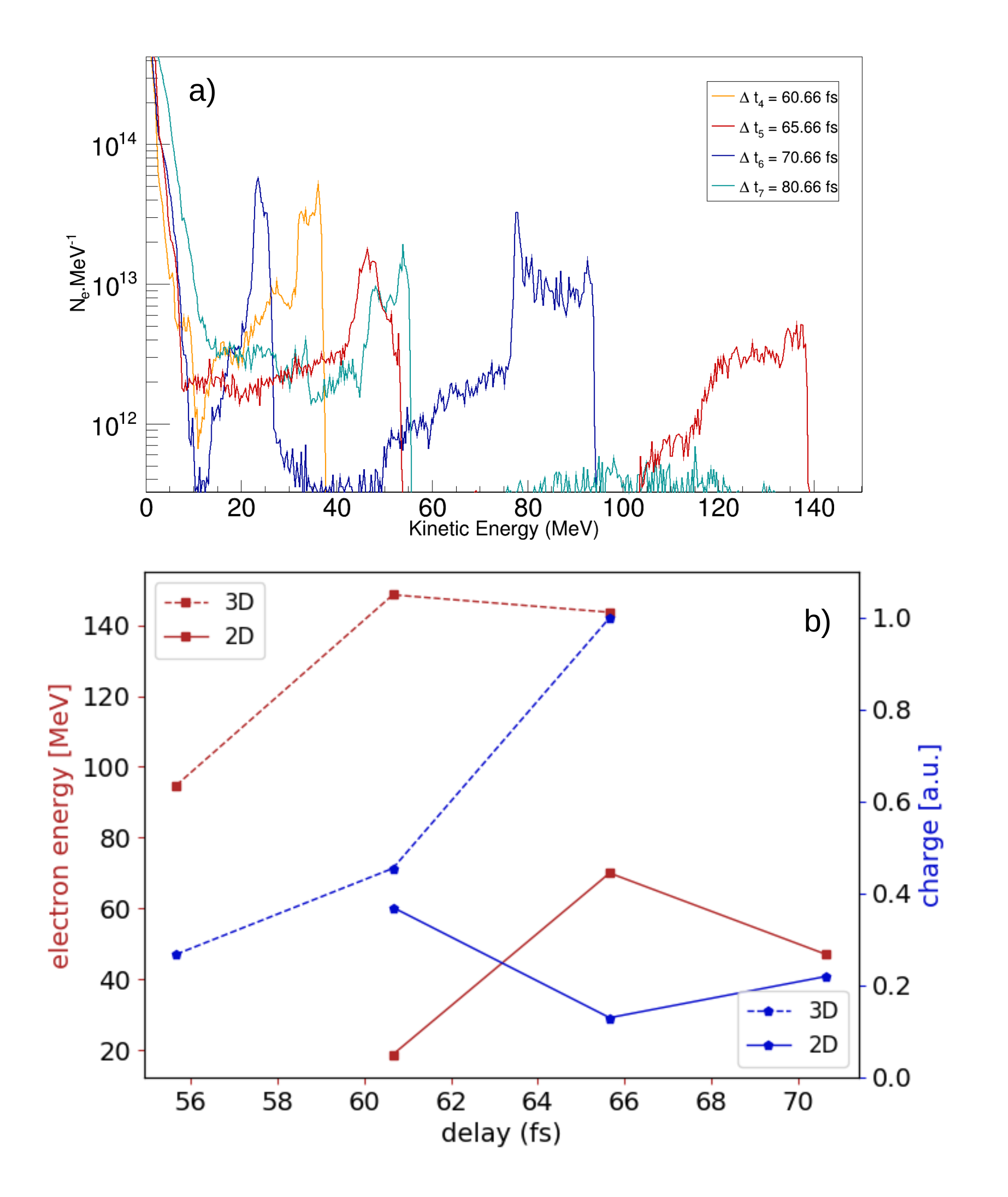}
	\caption{\label{scan}  a) 2D simulations comparing the energy spectra of accelerated electrons using the double-pulse scheme at $t=\SI{1300}{\femto\second}$, varying the relative delay between pulses where the size of the first cavity is $2r_b/c = \SI{65.66}{\femto\second}$, corresponding to normalized intensity of $a_{01} = 3.5$, for $(\gamma_p,a_0)=(30,5)$; b) robustness of double-pulse scheme in terms of delay adjustment, based on the energy and quantity of the injected electrons. The solid and dashed lines correspond the 2D and 3D simulations respectively. The preceding pulse for the 3D simulation has $a_{01} = 2.0$ rather than $a_{01} = 2.5$ for the 2D simulations.}

\end{figure}

Given the apparent advantages of the double-pulse scheme just demonstrated above, it is natural to ask under which conditions the injected electron beam properties are optimal. First, the ideal delay between the pulses would be expected to correspond to the size of the first cavity, or (following the notation used in Ref.~\citep{lu2007generating}) twice its longitudinal radius $2r_b\simeq 3.8a_{01}^{1/2}k_p^{-1}$. This is confirmed by a series of simulations with different delays but keeping the pulse energies equally divided, $\textit{i.e.}\, \mathrm{U}_2=\mathrm{U}_1$, and $a_{02}=a_{01}=3.5$---Fig.~\ref{scan}(a). Note that the scheme appears to be relatively robust with respect to a non-optimal pulse interval---Fig.~\ref{scan}(b). When choosing the optimum condition, we need to specify the desired electron bunch characteristics, and in this case we set the maximum achievable beam energy as the criterion. Based on these results, the maximum energy is attained if the center of the second laser pulse is delayed by $\Delta t=2r_b/c$ with respect to the first bubble. Within a tolerance of $\pm\SI{1.5}{\micro\meter}$ $(\pm\SI{5}{\femto\second})$, the scheme still works advantageously and delivers even higher flux albeit with lower maximum energy -- Fig.~\ref{scan}(b).

Repeating this exercise for the full 3D case, we see that the delay optimum is shifted to lower values probably because of stronger self-focusing/guiding effects and the shape of the wakefield in 3D, \textit{c.f.} Sec.~\ref{injthreshold}. On the other hand the slight anti-correlation of optimal charge and energy persists in 3D.

\subsection{Optimization of pulse energy division}\label{energyopt}

\begin{figure}[h]
	\includegraphics[scale=0.45]{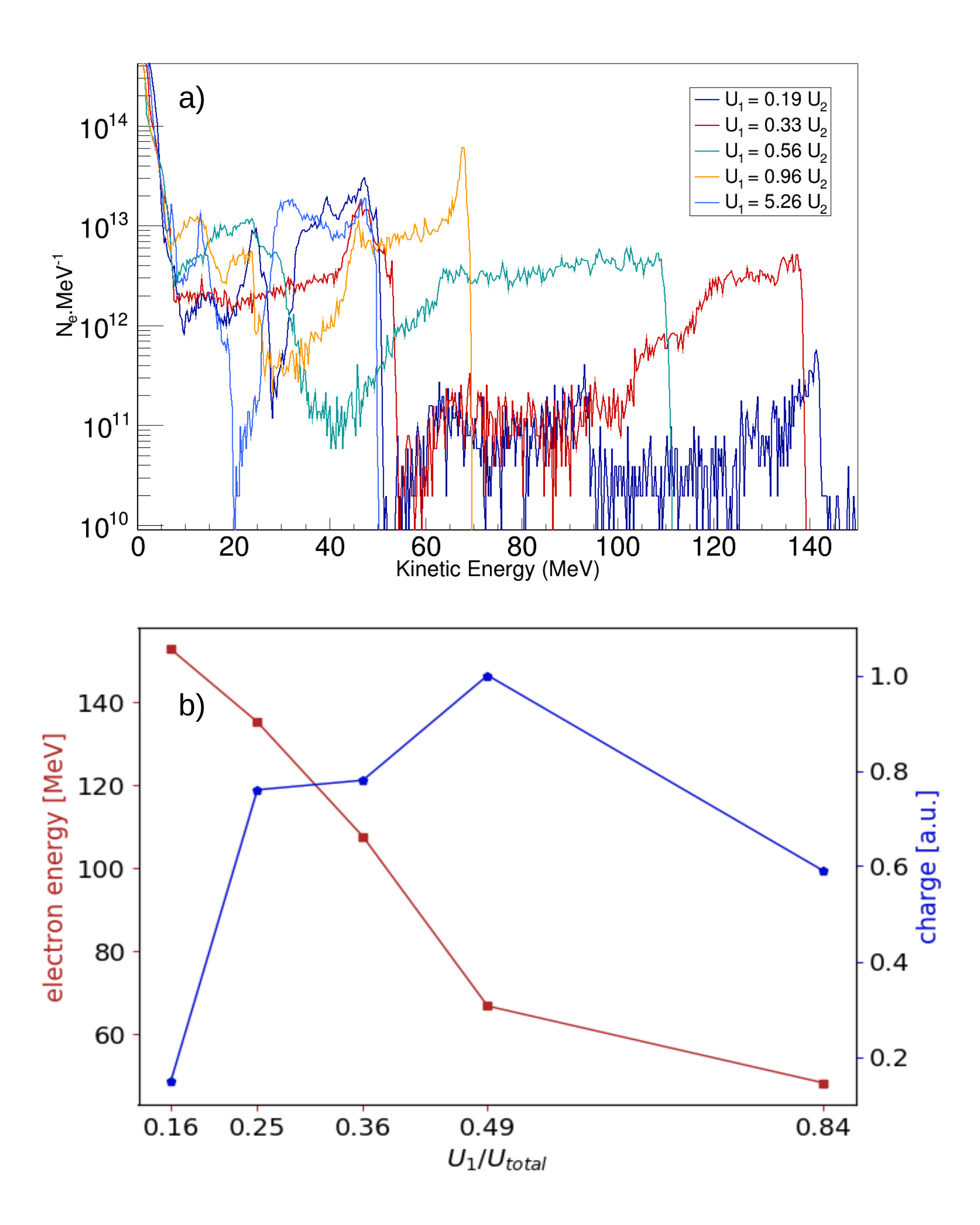}
	\caption{\label{2scan} a) 2D simulations comparing the energy spectra of accelerated electrons using the double-pulse scheme at $t=\SI{1300}{\femto\second}$ for different energy fraction of each pulse for $(\gamma_p,a_0)$ = (30,5). b) Scaling of injected charge (blue line) and energy (red line) with relative energy of the  first pulse.} 
\end{figure}

The optimal energy division was determined by performing a further series of five 2D simulations at the optimal pulse separation, dividing laser pulse energy between each pulses as follows: the original pulse amplitude is $a_0=5$. In the first simulation the first pulse carries a fraction of total pulse energy corresponding to the amplitude $a_{01}=2$, the remaining energy belongs to the second pulse. This was repeated in steps of $\Delta a_0=0.5$, up to the case where the first pulse carries an energy corresponding to $a_{01}=4$ and the remaining energy is allocated to the second pulse, which was placed on the rear side of the cavity created by the first pulse as discussed in Section~\ref{delayopt}. As a result, it is found for these specific laser parameters that the most efficient energy division is when the second pulse carries three times the energy of first pulse --- Fig.~\ref{2scan}(a,b). Generally there is a range of relative energy fraction where the injected beam can be optimized for energy or charge, or a combination thereof. For a total energy $U_{total}=\SI{3.4}{\joule}$ we take $U_1/U_{total}=0.25$, or $U_2/U_1=3$.

In order to find a more general rule for the energy fraction of each pulse, several other simulations are carried out with different laser intensity. The result is summarized in Table~\ref{frac}.

\begin{table}[ht]
\caption{\label{frac} Optimized values for energy fraction of each pulse. The pulse length, focal size and amplitude of the preceding pulse are kept the same in all simulations. Only the total energy of the pair of pulses and correspondingly the amplitude of the trailing pulse is changed. }
\begin{ruledtabular}
\begin{tabular}{llllll}
simulation $\#$ & $\textrm{U}_{\textrm{total}}[\SI{}{\joule}]$&\mbox{$\textrm{a}_{0,\textrm{total}}$}&\mbox{$a_{01}$}&\mbox{$a_{02}$}&\mbox{$U_2/U_1$}\\
\hline
			1 & 2.17 &4 & 2.5 & 3.12 & 1.6\\
			2 & 3.40  & 5 & 2.5 & 4.33 & 3\\
            3 & 4.89  & 6 & 2.5 & 5.45 & 4.9\\
\end{tabular}
\end{ruledtabular}
\end{table}
In all simulations, both pulses are the same in focal size and pulse length. It is evident that the intensity of the first pulse must be sufficiently high to meet the usual condition for the cavity formation, $a_{01}>2$, supplying an optimal quantity of ionized electrons for the second laser pulse. The rest of the laser energy can then be invested in the second laser pulse such that $a_0^2=a_{01}^2+a_{02}^2$, with $a_{02}/a_{01}>1$. Although these considerations for the pulse delay and amplitude ratio serve as a good initial guide to optimising the injection process, we will see later in Sec.~\ref{discussion} that propagation effects may complicate this choice.

\section{Self-injection threshold} \label{injthreshold}

\begin{figure}[h]
	\includegraphics[scale=2.2]{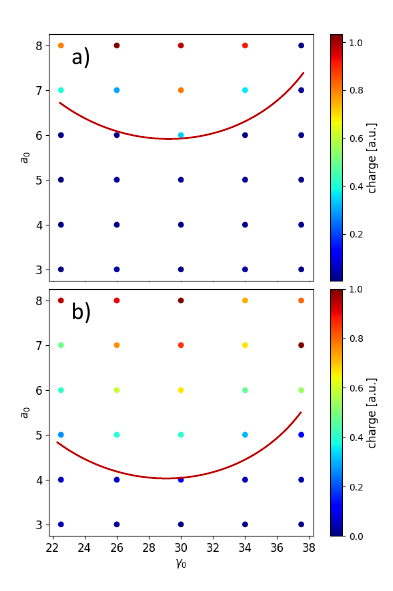}
	\caption{\label{chthr} Self-injection threshold for different laser amplitudes and phase velocities, where each point corresponds to a full 2D simulation. The color represents the relative amount of injected charge in a) the single-pulse and b) double-pulse schemes. The red solid line is injection threshold estimates based on our simulations with varying matching condition.  }
\end{figure}

The self-injection process depends both on the laser amplitude and wake phase velocity, which is tied to the group velocity of the laser pulse,~\cite{benedetti2013numerical} $(v_{gr} = c\sqrt{1-\omega_p^2/\omega^2})$, so to compare thresholds in the cavity regime for the double- and  single-pulse cases, a set of simulations were done with different laser amplitudes and target densities and then following Ref.~\citep{benedetti2013numerical} mapped in the $(\gamma_p,a_0)$ plane, where $\gamma_p$ is the Lorentz factor corresponding to the plasma wake moving with a velocity $v_p=v_{gr}$. The results are collected in Fig.~\ref{chthr}. 

\begin{figure}[h!]
%\begin{center}

	\includegraphics[scale=0.58]{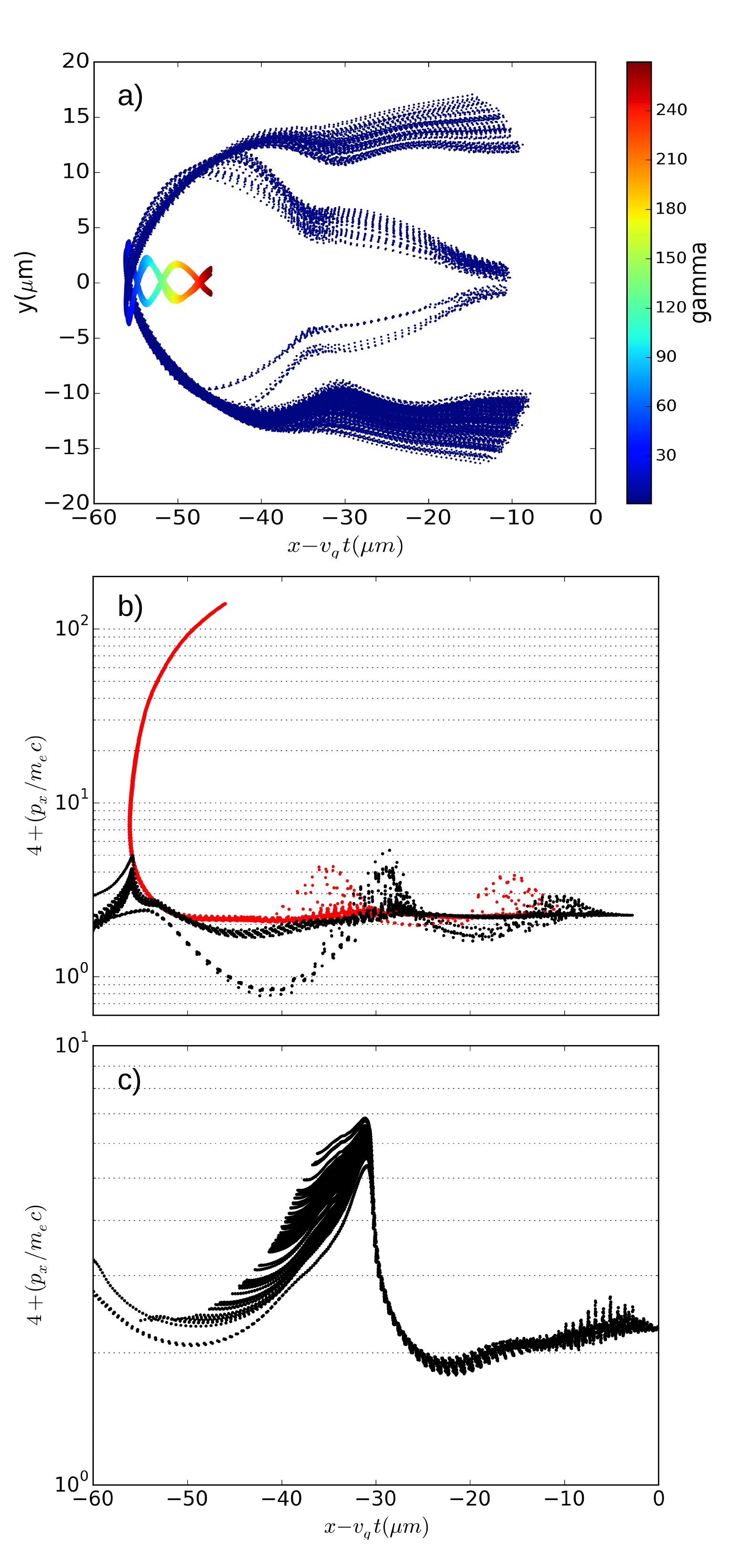}
	\caption{\label{trajectory} The electron trajectory for the 2D simulation of Fig.~\ref{numdens}, with $(\gamma_p,a_0)$ = (30,5); a) the trajectory of the injected electrons in the double-pulse scheme, b) the associated trajectory of injected and accelerated electrons in phase space for the double-pulse scheme. The red dots are the injected electrons and the black dots represent non-injected electrons that contribute to the plasma wave, c) the plasma orbit of the single-pulse scheme, in which there is no injection.  }
%	\end{center}
\end{figure}

The red solid lines in Fig.~\ref{chthr}(a) and (b) show the effective threshold when keeping the laser pulse length and waist (and therefore energy) the same while changing the target density (through $\gamma_p\simeq\omega_0/\omega_p$), for single- and double-pulse scheme respectively. Note that for this parameter scan we only ensure that the matching conditions  ($k_pw_0 =2\sqrt{a_0}$, $c\tau=\lambda_p/4$) are applied in some average sense but are not adjusted for each $(\gamma_p,a_0)$ pair; the procedure which would likely be followed in an experiment.  As one might expect, there is a density for which the wake phase velocity $(\gamma_p=30)$ seems to be optimal for electron injection: above and below this density, the cavity is driven non-resonantly with respect to injected bunch charge. Lower densities provide fewer electrons and correspondingly lower injected charge, and the wake has higher phase velocity. Therefore, it is more difficult for the electrons to become injected. Higher densities provide more charge but make the cavity regime more difficult to reach. 

Comparison of these charts between the single- and double-pulse schemes reveals a 30\% reduction of the threshold amplitude $a_0$ necessary for beam injection, corresponding to a $2\times$ reduction in (total) required intensity (or power), bringing obvious practical advantages.

To shed light on how a twin driver alters the injection process, we analyze trajectories of particles either side of the injection threshold. The latter has often been examined for the cavity regime before, but often with assumptions about the exact shape and (non-)rigidity of the cavity\cite{kalmykov2009electron,thomas2010scalings,Yi2011,Li2015}. 
Figure~\ref{trajectory} shows the trajectory of electrons in a 2D simulation for the double-pulse scheme Fig.~\ref{trajectory}(a) and (b), as well as the single-pulse scheme Fig.~\ref{trajectory}(c). 

 Notice how electrons from different lateral positions relative to the laser axis can still end up phase-synchronised after injection -- a feature studied in some detail in Ref.\citep{horny_ppcf2018}. Here, electrons are injected and accelerated up to $\SI{140}{\mega\electronvolt}$. The same injected particles are shown in ~\ref{trajectory}(b), which follows the electrons' trajectory in phase space, (red dots), together with the non-injected ones (black). Comparing this case to its single-pulse counterpart ~\ref{trajectory}(c), using the same total laser pulse energy, some of the  electrons which fail to get injected in the first cavity follow an extended orbit which deepens the potential well of the 2nd. In other words, when using two laser pulses the trapping separatrix of the 2nd cavity is modified in such a way that it becomes more favorable for electron injection, in an analogous fashion to an evolving bubble \cite{kalmykov2009electron}.

\begin{figure}[h]
\begin{center}
	\includegraphics[scale=2.3]{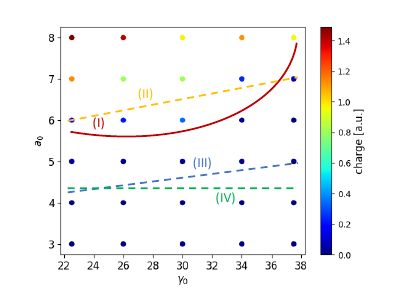}
	\caption{\label{chthr2} Self-injection threshold for different laser amplitudes and phase velocities, where each point corresponds to a full 2D simulation. The color represents the relative amount of injected charge in the single-pulse scheme, keeping the $c\tau_L = \lambda_p/4$ as the resonance condition for all points by adjusting the pulse duration and the focal spot size. The red (I) solid line is injection threshold estimates based on our simulations with fixed matching condition. The dashed green (IV), blue (III) and yellow (II) lines are the analogous thresholds according to Refs. ~\cite{thomas2010scalings,benedetti2013numerical,lu2007generating}, respectively.
	}
\end{center}

\end{figure}

In order to compare our single-pulse simulation results to previous theoretical work ~\cite{thomas2010scalings,benedetti2013numerical,lu2007generating}, we kept the point $\gamma_p=30$ as the optimum matching condition, where in fact  $c\tau_L = \lambda_p/4$. For each target density (or $\gamma_p$), the pulse duration and focal size were then adjusted to maintain a roughly spherical cavity shape --- Fig.~\ref{chthr2}.  The resulting red solid line (I) follows a similar trend to the dashed blue line (III)~\cite{benedetti2013numerical}, confirming that injection is easier in a higher density target, whereas higher thresholds can be expected with increasing wake phase velocity. The apparently higher laser amplitude threshold in our double-pulse case is likely because of the geometrical differences inherent in 2D and 3D simulation; a point we will examine in more detail shortly.

Several independent theoretical and numerical studies on self-injection threshold scaling have been previously published, but for various reasons it is difficult to compare these results quantitatively.  In our simulations we used a helium gas target with a short ramp of $\SI{12}{\micro\meter}$, and focused the laser $\SI{38}{\micro\meter}$ downstream of the ramp to ensure that the self-injection took place in the flat part of the density profile. Lu~\etal ~\cite{lu2007generating} used a plasma channel (dashed green line IV), but did not explicitly investigate the dependence of self-injection on $\gamma_p$: they found that self-injection occurs when the normalized blowout radius $k_pr_b \sim 4-5$. According to Ref.~\citep{thomas2010scalings} (dashed yellow line II) and Ref.~\citep{benedetti2013numerical} (dashed blue line III), the injection condition is dependant on the wake velocity (or $\gamma_p$); however Benedetti \etal\ predict a stronger dependency of self-injection on $\gamma_p$.  In Ref.~\citep{thomas2010scalings} a uniform electron density is assumed, whereas Benedetti \etal\ use a long target ramp with ionization enabled artificially only after a stable cavity is established. This deliberately suppresses the influence of the ramp on the injection process, allowing for a `clean' evaluation of the threshold. On the other hand such idealised conditions are probably difficult to realise experimentally.

In fact, it turns out that the injected beam current can be further increased by tuning the density ramp, as already demonstrated in Ref.~\citep{horny2018}. Our own 2D simulations to check the influence of the ramp size on electron injection at the point $(a_0,\gamma_p) = (5,30)$ in Fig.~\ref{chthr2} for ramp sizes from \SI{7}{\micro\meter} to \SI{22}{\micro\meter} predict an $17\%$ increase of injected charge over this range. These findings are not too sensitive to the placement of the laser focal spot as long as it is focused at least one cavity diameter beyond the top of the ramp. 

 It is well known that nonlinear laser pulse propagation, plasma wake evolution and injection can exhibit quantitative differences between 2D and 3D geometry~\cite{tsung2006simulation,davoine_pop2008}, so to get some idea of how geometrical effects might quantitatively alter our findings, several full 3D simulations were done at the points $(a_0,\gamma_p) = \{(4,26),(5,26),(3,30),(4,30),(5,30),(4,34),(5,34)\}$ and the results of single and double-pulse interactions compared in  Tab.~\ref{3D-bunch-quality}. Based on our 3D results, injection occurs for lower intensities for the double-pulse scheme and leads to a larger amount of injected charge, confirming the general trend seen in the 2D simulations. However, depending on the chosen delay the maximum energy is not necessarily higher for double-pulse scheme. This can be seen in the Tab.~\ref{3D-bunch-quality}; the delay for the cases with $\gamma_p = 30$ is chosen to be $\SI{66}{\femto\second}$, since for this density the delay was optimized---Fig.~\ref{scan}(b). In this case both energy and charge are higher than in the corresponding single-pulse case. For other target densities, $\gamma_p = 26,34$, the delay was not optimized and this may be the reason for the lower electron energy observed; however, the amount of injected charge is still higher than the single-pulse scheme. Finally, we observe that the emittance in the double-pulse scheme is generally comparable to the value from the single-pulse scheme despite higher injected charge and in some cases higher beam energy for the cases examined. The energy could probably be increased further by devising a proper matching condition for double-pulse drive.

 \begin{table*}
\caption{\label{3D-bunch-quality} 3-dimensional simulation results comparing the injected electron bunch properties (charge, energy and emittance) of single- and double-pulse schemes over one Rayleigh length. The first column indicates the point in the density-amplitude plane of Fig.~\ref{chthr2}. The emittance is calculated according to $\epsilon_{N,rms}=\frac{\overline{p}_x}{m_0c}\sqrt{\langle y^2 \rangle \langle {y^\prime}^2 \rangle - {\langle y y^\prime \rangle}^2}$, where $\overline{p}_x$ denotes the average longitudinal momentum, and $y^\prime = \frac{p_y}{px} = \tan\alpha$ is the transverse angle with respect to the ideal trajectory.}
\begin{ruledtabular}
\begin{tabular}{ccccccc}
 &\multicolumn{2}{c}{Charge (\SI{}{\pico\coulomb})}&\multicolumn{2}{c}{max. Energy (\SI{}{\mega\electronvolt})}&\multicolumn{2}{c}{Normalized emittance (\SI{}{\milli\meter\milli\radian})}\\
 $(\gamma_p,a_0)$&single-pulse&double-pulse&single-pulse
&double-pulse&single-pulse&double-pulse\\ \hline
 (26,4) & 131.02 & 143.35 & 211.86 & 172.62 & 5.04 & 8.94\\
 (26,5) & 219.98 & 346.61 & 289.63 & 194.74 & 4.59 & 5.98\\
 (30,3) &3.24 & 131 & 138.23 & 143.66 & 9.67& 5.72\\
 (30,4) & 122 & 209 & 197.51 & 201.84 & 6.19 & 3.21\\
 (30,5) & 226 & 357 & 222.36 & 238.52 & 7.60 & 3.80\\
 (34,4) & 6.54 & 176.17 & 187.93 & 144.79 & 2.50 & 4.47\\
 (34,5) & 179.6 & 338.29 & 236.75 & 141.45 & 4.89 & 2.97\\
\end{tabular}
\end{ruledtabular}
\end{table*}

 \begin{figure}[]
\begin{center}

	\includegraphics[scale=0.82]{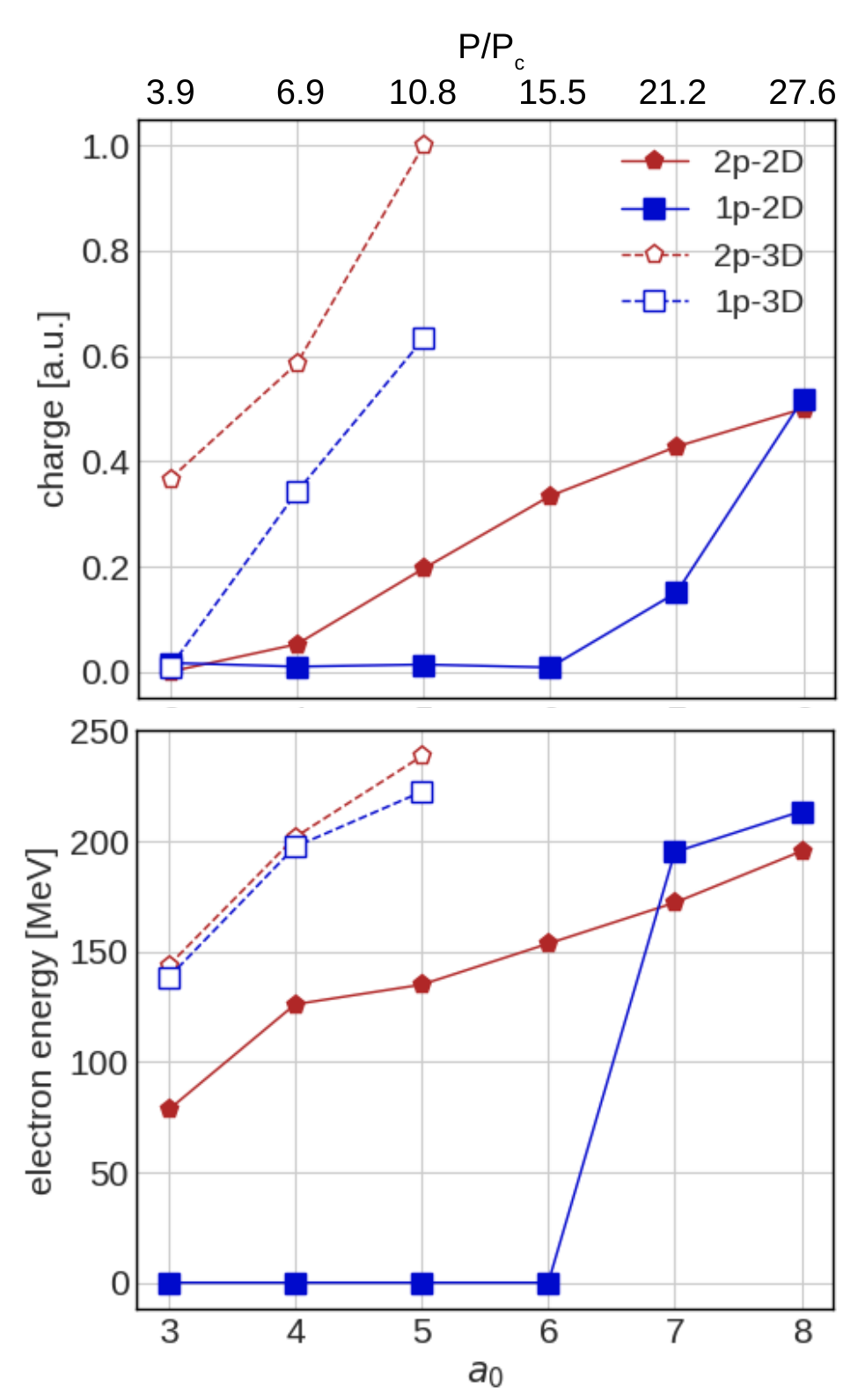}
	\caption{\label{2d3d-result} Comparison between single- (1p) and double-pulse (2p) for the same simulation parameters in 2D (solid lines) and 3D (dashed lines); The 3D results confirm the advantage of double-pulse over the single-pulse scheme in terms of the amount of injected charge and the maximum energy. In general, 3D simulations yield higher amount of charge than the 2D simulations. The upper axis shows the $P/P_c$ for each of the simulations, where the spot size is $w_0 = \SI{10}{\micro\meter}$ and the $\gamma_p = 30$.}
	\end{center}
\end{figure}
 \begin{figure*}[]
	\includegraphics[scale=0.6]{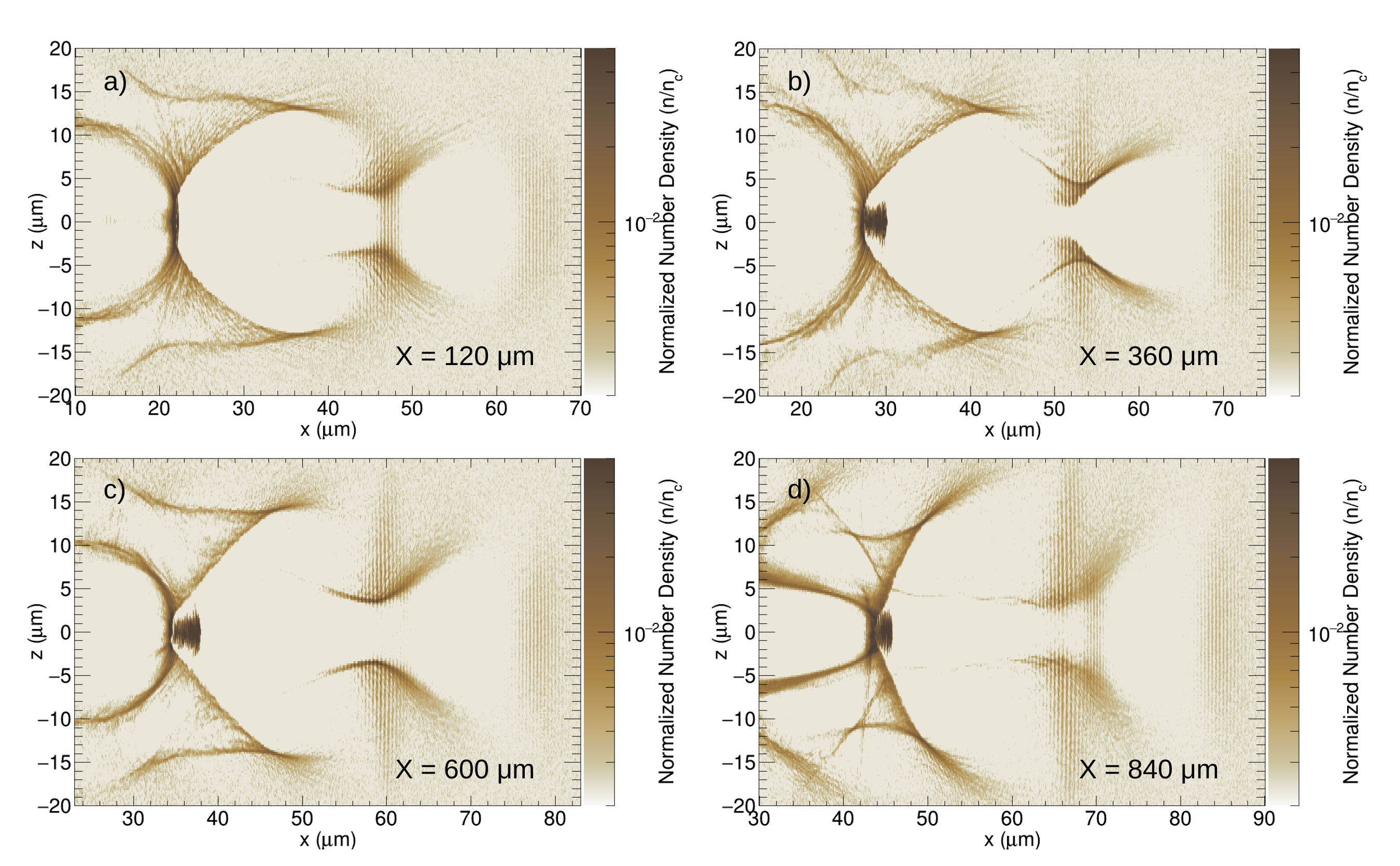}
	\caption{ Electron number density plot of a 3-dimensional simulation for $(\gamma_p,a_0) = (30,5)$. The simulation is carried out for twice a Rayleigh length ($\SI{2.8}{\pico\second}$), to check the stability of the beam and matching condition. }
	\label{3D-long} 

\end{figure*}

 As expected, the same general behaviour as in Fig.~\ref{chthr} is observed for the 3D simulations, where there is an optimal density ($\gamma_p = 30$) for injection. Furthermore, the higher injection threshold seen in our 2D simulations in Fig.~\ref{chthr2} compared to other works can also be accounted for by the geometry. This difference is displayed more explicitly in  Fig.~\ref{2d3d-result}, which shows how the beam charge and energy increases with laser power at fixed plasma density (here $\gamma_p = 30$, or $n_e=\SI{1.9E18}{\per\cubic\centi\meter}$). According to Fig.~\ref{2d3d-result}a), the single-pulse threshold $ P/P_c > 5$ previously observed in a number of experiments\cite{mangles2012self,froula2009measurements} is effectively reduced to a value $P/P_c \sim 2-3$ for the double pulse scheme.

\section{Discussion and conclusion} \label{discussion}
We have shown that a tandem-pulse wakefield accelerator in the nonlinear regime can offer significant advantages over the conventional single-pulse driver, yielding higher electron currents and energies thanks to an enhanced cavity size and a lower electron injection threshold for the same total laser energy. Full 3D simulations show that the injection threshold intensity or power can be reduced by at least a factor of two, confirming a wider 2D parameter scan for a set of laser intensities and target densities. 

The latter simulations indicate that further optimisation of this scheme may be possible by adjusting the pulse separation and relative amplitudes to allow for dynamical evolution of the cavity over longer propagation distances. To explore this properly, a comprehensive 3D or quasi-3D~\cite{benedetti2013numerical} set of simulations would be needed in a multi-dimensional parameter space; a task beyond the scope of the present work. 

Nevertheless, it is of interest to examine how robust the tandem scheme is over significantly longer propagation distances, so a proof-of-principle 3D simulation corresponding to two Rayleigh lengths has been carried out for $(\gamma_p,a_0) = (30,5)$. A time series of the electron number density is plotted in Fig.~\ref{3D-long}, in which the laser pulse evolution is also partially visible. Starting with the target density corresponding to $\gamma_p = 30$, the cavity has a spherical shape at the beginning, Fig.~\ref{3D-long}-(a). For the chosen laser parameter with focal spot size of $w_0 = \SI{10}{\micro\meter}$,  two Rayleigh lengths corresponds to $\SI{785}{\micro\meter}$, so the laser intensity should undergo a significant decrease unless there is some self-guiding \cite{lu2007generating}. Moreover, because of the aggregation of electrons at the back of the first cavity and change of the target density, the second laser pulse may also be affected by additional refraction, which causes the cavity size to decrease in the longitudinal direction. The latter effect eventually leads to loss of trapped electrons from the back of the second cavity in this case. 

Despite these factors, acceleration of $\SI{0.29}{\nano\coulomb}$ electron bunches up to $\SI{536}{\mega\electronvolt}$ with a peak at $\SI{395}{\mega\electronvolt}$ and energy spread $\Delta E/E = 12.7\%$ at FWHM within this distance is observed.  
By comparison, the matched single-pulse scheme yields a beam with $\SI{0.22}{\nano\coulomb}$ charge and maximum energy $\SI{450}{\mega\electronvolt}$ with FWHM bandwidth $\Delta E/E = 33 \%$ and a peak at $\SI{300}{\mega\electronvolt}$. On the other hand, the cavity shape remains roughly spherical over this distance and exhibits no leakage of the trapped beam, raising the possibility of stabilizing the tandem-pulse propagation over many Rayleigh lengths in the blowout regime via a modified matching condition. For example, more control over the laser pulse evolution might be achieved by starting with a slightly mis-matched configuration such that the 2nd cavity is initially elongated compared to the first, or by setting a longer-than-optimal delay suggested by Fig.\ref{scan}(b).  

Clearly further study is needed to mitigate and properly exploit the interplay between relativistic propagation effects and dynamical refraction to achieve a fully `matched' tandem-pulse scheme. At this point we can conclude that this scheme has tangible advantages for producing electron beams with modest energies (10s to 100s of MeV) suited to x-ray generation with smaller TW lasers, perhaps with high repetition rate. Additional trial simulations with $\sim \SI{100}{\milli\joule}$ laser pulse energy, $\SI{10}{\femto\second}$ pulse duration and focal spot size down to $\SI{5}{\micro\meter}$ confirm the advantage of double-pulse drive over single-pulse.
Finally, it is possible that trains of 3 or more pulses might permit even greater long-term control over cavity dynamics and beam energies for the same total pulse energy, extending the original 1D pulse-train concept \cite{umstadter1994nonlinear} to the nonlinear, three-dimensional regime.

\begin{acknowledgments}
This work was carried out in the framework of the \textit{Ju}SPARC project (J\"ulich Short Pulsed Particle and Radiation Centre) at Forschungszentrum J\"ulich. The authors acknowledge computing time grants from the J\"ulich Supercomputer Centre, projects JPGI61 and JZAM04. One of the authors (ZMC) gratefully acknowledges fruitful discussions with Ilhan Engin and Andr\'e Sobotta as well as visualization support by Jens Henrik G\"obbert. 
\end{acknowledgments}

\nocite{*}
\bibliography{mainmain}% Produces the bibliography via BibTeX.

\begin{thebibliography}{58}%
\makeatletter
\providecommand \@ifxundefined [1]{%
 \@ifx{#1\undefined}
}%
\providecommand \@ifnum [1]{%
 \ifnum #1\expandafter \@firstoftwo
 \else \expandafter \@secondoftwo
 \fi
}%
\providecommand \@ifx [1]{%
 \ifx #1\expandafter \@firstoftwo
 \else \expandafter \@secondoftwo
 \fi
}%
\providecommand \natexlab [1]{#1}%
\providecommand \enquote  [1]{``#1''}%
\providecommand \bibnamefont  [1]{#1}%
\providecommand \bibfnamefont [1]{#1}%
\providecommand \citenamefont [1]{#1}%
\providecommand \href@noop [0]{\@secondoftwo}%
\providecommand \href [0]{\begingroup \@sanitize@url \@href}%
\providecommand \@href[1]{\@@startlink{#1}\@@href}%
\providecommand \@@href[1]{\endgroup#1\@@endlink}%
\providecommand \@sanitize@url [0]{\catcode `\\12\catcode `\$12\catcode
  `\&12\catcode `\#12\catcode `\^12\catcode `\_12\catcode `\%12\relax}%
\providecommand \@@startlink[1]{}%
\providecommand \@@endlink[0]{}%
\providecommand \url  [0]{\begingroup\@sanitize@url \@url }%
\providecommand \@url [1]{\endgroup\@href {#1}{\urlprefix }}%
\providecommand \urlprefix  [0]{URL }%
\providecommand \Eprint [0]{\href }%
\providecommand \doibase [0]{http://dx.doi.org/}%
\providecommand \selectlanguage [0]{\@gobble}%
\providecommand \bibinfo  [0]{\@secondoftwo}%
\providecommand \bibfield  [0]{\@secondoftwo}%
\providecommand \translation [1]{[#1]}%
\providecommand \BibitemOpen [0]{}%
\providecommand \bibitemStop [0]{}%
\providecommand \bibitemNoStop [0]{.\EOS\space}%
\providecommand \EOS [0]{\spacefactor3000\relax}%
\providecommand \BibitemShut  [1]{\csname bibitem#1\endcsname}%
\let\auto@bib@innerbib\@empty
%</preamble>
\bibitem [{\citenamefont {Malka}(2017)}]{malka_cern2016}%
  \BibitemOpen
  \bibfield  {author} {\bibinfo {author} {\bibfnamefont {V.}~\bibnamefont
  {Malka}},\ }\bibfield  {title} {\enquote {\bibinfo {title} {{Plasma Wake
  Accelerators: Introduction and Historical Overview}},}\ }\href {\doibase
  10.5170/CERN-2016-001.1} {\bibfield  {journal} {\bibinfo  {journal} {CERN
  Yellow Reports}\ }\textbf {\bibinfo {volume} {001}},\ \bibinfo {pages}
  {23--29} (\bibinfo {year} {2017})}\BibitemShut {NoStop}%
\bibitem [{\citenamefont {Lemke}\ \emph {et~al.}(2013)\citenamefont {Lemke},
  \citenamefont {Bressler}, \citenamefont {Chen}, \citenamefont {Fritz},
  \citenamefont {Gaffney}, \citenamefont {Galler}, \citenamefont {Gawelda},
  \citenamefont {Haldrup}, \citenamefont {Hartsock}, \citenamefont {Ihee} \emph
  {et~al.}}]{lemke2013femtosecond}%
  \BibitemOpen
  \bibfield  {author} {\bibinfo {author} {\bibfnamefont {H.~T.}\ \bibnamefont
  {Lemke}}, \bibinfo {author} {\bibfnamefont {C.}~\bibnamefont {Bressler}},
  \bibinfo {author} {\bibfnamefont {L.~X.}\ \bibnamefont {Chen}}, \bibinfo
  {author} {\bibfnamefont {D.~M.}\ \bibnamefont {Fritz}}, \bibinfo {author}
  {\bibfnamefont {K.~J.}\ \bibnamefont {Gaffney}}, \bibinfo {author}
  {\bibfnamefont {A.}~\bibnamefont {Galler}}, \bibinfo {author} {\bibfnamefont
  {W.}~\bibnamefont {Gawelda}}, \bibinfo {author} {\bibfnamefont
  {K.}~\bibnamefont {Haldrup}}, \bibinfo {author} {\bibfnamefont {R.~W.}\
  \bibnamefont {Hartsock}}, \bibinfo {author} {\bibfnamefont {H.}~\bibnamefont
  {Ihee}},  \emph {et~al.},\ }\href@noop {} {\bibfield  {journal} {\bibinfo
  {journal} {The Journal of Physical Chemistry A}\ }\textbf {\bibinfo {volume}
  {117}},\ \bibinfo {pages} {735--740} (\bibinfo {year} {2013})}\BibitemShut
  {NoStop}%
\bibitem [{\citenamefont {Rousse}, \citenamefont {Rischel},\ and\ \citenamefont
  {Gauthier}(2001)}]{rousse2001femtosecond}%
  \BibitemOpen
  \bibfield  {author} {\bibinfo {author} {\bibfnamefont {A.}~\bibnamefont
  {Rousse}}, \bibinfo {author} {\bibfnamefont {C.}~\bibnamefont {Rischel}}, \
  and\ \bibinfo {author} {\bibfnamefont {J.-C.}\ \bibnamefont {Gauthier}},\
  }\href@noop {} {\bibfield  {journal} {\bibinfo  {journal} {Reviews of Modern
  Physics}\ }\textbf {\bibinfo {volume} {73}},\ \bibinfo {pages} {17} (\bibinfo
  {year} {2001})}\BibitemShut {NoStop}%
\bibitem [{\citenamefont {Uesaka}\ \emph {et~al.}(2000)\citenamefont {Uesaka},
  \citenamefont {Kotaki}, \citenamefont {Nakajima}, \citenamefont {Harano},
  \citenamefont {Kinoshita}, \citenamefont {Watanabe}, \citenamefont {Ueda},
  \citenamefont {Yoshii}, \citenamefont {Kando}, \citenamefont {Dewa} \emph
  {et~al.}}]{uesaka2000generation}%
  \BibitemOpen
  \bibfield  {author} {\bibinfo {author} {\bibfnamefont {M.}~\bibnamefont
  {Uesaka}}, \bibinfo {author} {\bibfnamefont {H.}~\bibnamefont {Kotaki}},
  \bibinfo {author} {\bibfnamefont {K.}~\bibnamefont {Nakajima}}, \bibinfo
  {author} {\bibfnamefont {H.}~\bibnamefont {Harano}}, \bibinfo {author}
  {\bibfnamefont {K.}~\bibnamefont {Kinoshita}}, \bibinfo {author}
  {\bibfnamefont {T.}~\bibnamefont {Watanabe}}, \bibinfo {author}
  {\bibfnamefont {T.}~\bibnamefont {Ueda}}, \bibinfo {author} {\bibfnamefont
  {K.}~\bibnamefont {Yoshii}}, \bibinfo {author} {\bibfnamefont
  {M.}~\bibnamefont {Kando}}, \bibinfo {author} {\bibfnamefont
  {H.}~\bibnamefont {Dewa}},  \emph {et~al.},\ }\href@noop {} {\bibfield
  {journal} {\bibinfo  {journal} {Nuclear Instruments and Methods in Physics
  Research Section A: Accelerators, Spectrometers, Detectors and Associated
  Equipment}\ }\textbf {\bibinfo {volume} {455}},\ \bibinfo {pages} {90--98}
  (\bibinfo {year} {2000})}\BibitemShut {NoStop}%
\bibitem [{\citenamefont {Rousse}\ \emph {et~al.}(2004)\citenamefont {Rousse},
  \citenamefont {Phuoc}, \citenamefont {Shah}, \citenamefont {Pukhov},
  \citenamefont {Lefebvre}, \citenamefont {Malka}, \citenamefont {Kiselev},
  \citenamefont {Burgy}, \citenamefont {Rousseau}, \citenamefont {Umstadter}
  \emph {et~al.}}]{rousse2004production}%
  \BibitemOpen
  \bibfield  {author} {\bibinfo {author} {\bibfnamefont {A.}~\bibnamefont
  {Rousse}}, \bibinfo {author} {\bibfnamefont {K.~T.}\ \bibnamefont {Phuoc}},
  \bibinfo {author} {\bibfnamefont {R.}~\bibnamefont {Shah}}, \bibinfo {author}
  {\bibfnamefont {A.}~\bibnamefont {Pukhov}}, \bibinfo {author} {\bibfnamefont
  {E.}~\bibnamefont {Lefebvre}}, \bibinfo {author} {\bibfnamefont
  {V.}~\bibnamefont {Malka}}, \bibinfo {author} {\bibfnamefont
  {S.}~\bibnamefont {Kiselev}}, \bibinfo {author} {\bibfnamefont
  {F.}~\bibnamefont {Burgy}}, \bibinfo {author} {\bibfnamefont {J.-P.}\
  \bibnamefont {Rousseau}}, \bibinfo {author} {\bibfnamefont {D.}~\bibnamefont
  {Umstadter}},  \emph {et~al.},\ }\href@noop {} {\bibfield  {journal}
  {\bibinfo  {journal} {Physical Review Letters}\ }\textbf {\bibinfo {volume}
  {93}},\ \bibinfo {pages} {135005} (\bibinfo {year} {2004})}\BibitemShut
  {NoStop}%
\bibitem [{\citenamefont {Kneip}\ \emph {et~al.}(2008)\citenamefont {Kneip},
  \citenamefont {Nagel}, \citenamefont {Bellei}, \citenamefont {Bourgeois},
  \citenamefont {Dangor}, \citenamefont {Gopal}, \citenamefont {Heathcote},
  \citenamefont {Mangles}, \citenamefont {Marques}, \citenamefont {Maksimchuk}
  \emph {et~al.}}]{kneip2008observation}%
  \BibitemOpen
  \bibfield  {author} {\bibinfo {author} {\bibfnamefont {S.}~\bibnamefont
  {Kneip}}, \bibinfo {author} {\bibfnamefont {S.}~\bibnamefont {Nagel}},
  \bibinfo {author} {\bibfnamefont {C.}~\bibnamefont {Bellei}}, \bibinfo
  {author} {\bibfnamefont {N.}~\bibnamefont {Bourgeois}}, \bibinfo {author}
  {\bibfnamefont {A.}~\bibnamefont {Dangor}}, \bibinfo {author} {\bibfnamefont
  {A.}~\bibnamefont {Gopal}}, \bibinfo {author} {\bibfnamefont
  {R.}~\bibnamefont {Heathcote}}, \bibinfo {author} {\bibfnamefont
  {S.}~\bibnamefont {Mangles}}, \bibinfo {author} {\bibfnamefont
  {J.}~\bibnamefont {Marques}}, \bibinfo {author} {\bibfnamefont
  {A.}~\bibnamefont {Maksimchuk}},  \emph {et~al.},\ }\href@noop {} {\bibfield
  {journal} {\bibinfo  {journal} {Physical Review Letters}\ }\textbf {\bibinfo
  {volume} {100}},\ \bibinfo {pages} {105006} (\bibinfo {year}
  {2008})}\BibitemShut {NoStop}%
\bibitem [{\citenamefont {Kneip}\ \emph {et~al.}(2010)\citenamefont {Kneip},
  \citenamefont {McGuffey}, \citenamefont {Martins}, \citenamefont {Martins},
  \citenamefont {Bellei}, \citenamefont {Chvykov}, \citenamefont {Dollar},
  \citenamefont {Fonseca}, \citenamefont {Huntington}, \citenamefont
  {Kalintchenko} \emph {et~al.}}]{kneip2010bright}%
  \BibitemOpen
  \bibfield  {author} {\bibinfo {author} {\bibfnamefont {S.}~\bibnamefont
  {Kneip}}, \bibinfo {author} {\bibfnamefont {C.}~\bibnamefont {McGuffey}},
  \bibinfo {author} {\bibfnamefont {J.}~\bibnamefont {Martins}}, \bibinfo
  {author} {\bibfnamefont {S.}~\bibnamefont {Martins}}, \bibinfo {author}
  {\bibfnamefont {C.}~\bibnamefont {Bellei}}, \bibinfo {author} {\bibfnamefont
  {V.}~\bibnamefont {Chvykov}}, \bibinfo {author} {\bibfnamefont
  {F.}~\bibnamefont {Dollar}}, \bibinfo {author} {\bibfnamefont
  {R.}~\bibnamefont {Fonseca}}, \bibinfo {author} {\bibfnamefont
  {C.}~\bibnamefont {Huntington}}, \bibinfo {author} {\bibfnamefont
  {G.}~\bibnamefont {Kalintchenko}},  \emph {et~al.},\ }\href@noop {}
  {\bibfield  {journal} {\bibinfo  {journal} {Nature Physics}\ }\textbf
  {\bibinfo {volume} {6}},\ \bibinfo {pages} {980} (\bibinfo {year}
  {2010})}\BibitemShut {NoStop}%
\bibitem [{\citenamefont {Corde}\ \emph {et~al.}(2011)\citenamefont {Corde},
  \citenamefont {Phuoc}, \citenamefont {Fitour}, \citenamefont {Faure},
  \citenamefont {Tafzi}, \citenamefont {Goddet}, \citenamefont {Malka},\ and\
  \citenamefont {Rousse}}]{corde2011controlled}%
  \BibitemOpen
  \bibfield  {author} {\bibinfo {author} {\bibfnamefont {S.}~\bibnamefont
  {Corde}}, \bibinfo {author} {\bibfnamefont {K.~T.}\ \bibnamefont {Phuoc}},
  \bibinfo {author} {\bibfnamefont {R.}~\bibnamefont {Fitour}}, \bibinfo
  {author} {\bibfnamefont {J.}~\bibnamefont {Faure}}, \bibinfo {author}
  {\bibfnamefont {A.}~\bibnamefont {Tafzi}}, \bibinfo {author} {\bibfnamefont
  {J.-P.}\ \bibnamefont {Goddet}}, \bibinfo {author} {\bibfnamefont
  {V.}~\bibnamefont {Malka}}, \ and\ \bibinfo {author} {\bibfnamefont
  {A.}~\bibnamefont {Rousse}},\ }\href@noop {} {\bibfield  {journal} {\bibinfo
  {journal} {Physical Review Letters}\ }\textbf {\bibinfo {volume} {107}},\
  \bibinfo {pages} {255003} (\bibinfo {year} {2011})}\BibitemShut {NoStop}%
\bibitem [{\citenamefont {Tajima}\ and\ \citenamefont
  {Dawson}(1979)}]{tajima1979laser}%
  \BibitemOpen
  \bibfield  {author} {\bibinfo {author} {\bibfnamefont {T.}~\bibnamefont
  {Tajima}}\ and\ \bibinfo {author} {\bibfnamefont {J.~M.}\ \bibnamefont
  {Dawson}},\ }\href@noop {} {\bibfield  {journal} {\bibinfo  {journal}
  {Physical Review Letters}\ }\textbf {\bibinfo {volume} {43}},\ \bibinfo
  {pages} {267} (\bibinfo {year} {1979})}\BibitemShut {NoStop}%
\bibitem [{\citenamefont {Umstadter}, \citenamefont {Esarey},\ and\
  \citenamefont {Kim}(1994)}]{umstadter1994nonlinear}%
  \BibitemOpen
  \bibfield  {author} {\bibinfo {author} {\bibfnamefont {D.}~\bibnamefont
  {Umstadter}}, \bibinfo {author} {\bibfnamefont {E.}~\bibnamefont {Esarey}}, \
  and\ \bibinfo {author} {\bibfnamefont {J.}~\bibnamefont {Kim}},\ }\href@noop
  {} {\bibfield  {journal} {\bibinfo  {journal} {Physical Review Letters}\
  }\textbf {\bibinfo {volume} {72}},\ \bibinfo {pages} {1224} (\bibinfo {year}
  {1994})}\BibitemShut {NoStop}%
\bibitem [{\citenamefont {Dalla}\ and\ \citenamefont
  {Lontano}(1994)}]{dalla1994large}%
  \BibitemOpen
  \bibfield  {author} {\bibinfo {author} {\bibfnamefont {S.}~\bibnamefont
  {Dalla}}\ and\ \bibinfo {author} {\bibfnamefont {M.}~\bibnamefont
  {Lontano}},\ }\href@noop {} {\bibfield  {journal} {\bibinfo  {journal}
  {Physical Review E}\ }\textbf {\bibinfo {volume} {49}},\ \bibinfo {pages}
  {R1819} (\bibinfo {year} {1994})}\BibitemShut {NoStop}%
\bibitem [{\citenamefont {Kim}\ \emph {et~al.}(2007)\citenamefont {Kim},
  \citenamefont {Kim}, \citenamefont {Kim}, \citenamefont {Ko},\ and\
  \citenamefont {Suk}}]{kim2007double}%
  \BibitemOpen
  \bibfield  {author} {\bibinfo {author} {\bibfnamefont {C.}~\bibnamefont
  {Kim}}, \bibinfo {author} {\bibfnamefont {J.-C.~B.}\ \bibnamefont {Kim}},
  \bibinfo {author} {\bibfnamefont {K.}~\bibnamefont {Kim}}, \bibinfo {author}
  {\bibfnamefont {I.~S.}\ \bibnamefont {Ko}}, \ and\ \bibinfo {author}
  {\bibfnamefont {H.}~\bibnamefont {Suk}},\ }\bibfield  {title} {\enquote
  {\bibinfo {title} {Double pulse laser wakefield accelerator},}\ }\href@noop
  {} {\bibfield  {journal} {\bibinfo  {journal} {Physics Letters A}\ }\textbf
  {\bibinfo {volume} {370}},\ \bibinfo {pages} {310--315} (\bibinfo {year}
  {2007})}\BibitemShut {NoStop}%
\bibitem [{\citenamefont {Hooker}\ \emph {et~al.}(2014)\citenamefont {Hooker},
  \citenamefont {Bartolini}, \citenamefont {Mangles}, \citenamefont
  {T{\"u}nnermann}, \citenamefont {Corner}, \citenamefont {Limpert},
  \citenamefont {Seryi},\ and\ \citenamefont {Walczak}}]{hooker2014multi}%
  \BibitemOpen
  \bibfield  {author} {\bibinfo {author} {\bibfnamefont {S.}~\bibnamefont
  {Hooker}}, \bibinfo {author} {\bibfnamefont {R.}~\bibnamefont {Bartolini}},
  \bibinfo {author} {\bibfnamefont {S.}~\bibnamefont {Mangles}}, \bibinfo
  {author} {\bibfnamefont {A.}~\bibnamefont {T{\"u}nnermann}}, \bibinfo
  {author} {\bibfnamefont {L.}~\bibnamefont {Corner}}, \bibinfo {author}
  {\bibfnamefont {J.}~\bibnamefont {Limpert}}, \bibinfo {author} {\bibfnamefont
  {A.}~\bibnamefont {Seryi}}, \ and\ \bibinfo {author} {\bibfnamefont
  {R.}~\bibnamefont {Walczak}},\ }\href@noop {} {\bibfield  {journal} {\bibinfo
   {journal} {Journal of Physics B: Atomic, Molecular and Optical Physics}\
  }\textbf {\bibinfo {volume} {47}},\ \bibinfo {pages} {234003} (\bibinfo
  {year} {2014})}\BibitemShut {NoStop}%
\bibitem [{\citenamefont {Bonnaud}, \citenamefont {Teychenn{\'e}},\ and\
  \citenamefont {Bobin}(1994)}]{bonnaud1994wake}%
  \BibitemOpen
  \bibfield  {author} {\bibinfo {author} {\bibfnamefont {G.}~\bibnamefont
  {Bonnaud}}, \bibinfo {author} {\bibfnamefont {D.}~\bibnamefont
  {Teychenn{\'e}}}, \ and\ \bibinfo {author} {\bibfnamefont {J.-L.}\
  \bibnamefont {Bobin}},\ }\href@noop {} {\bibfield  {journal} {\bibinfo
  {journal} {Physical Review E}\ }\textbf {\bibinfo {volume} {50}},\ \bibinfo
  {pages} {R36} (\bibinfo {year} {1994})}\BibitemShut {NoStop}%
\bibitem [{\citenamefont {Tomassini}\ \emph {et~al.}(2017)\citenamefont
  {Tomassini}, \citenamefont {De~Nicola}, \citenamefont {Labate}, \citenamefont
  {Londrillo}, \citenamefont {Fedele}, \citenamefont {Terzani},\ and\
  \citenamefont {Gizzi}}]{tomassini2017resonant}%
  \BibitemOpen
  \bibfield  {author} {\bibinfo {author} {\bibfnamefont {P.}~\bibnamefont
  {Tomassini}}, \bibinfo {author} {\bibfnamefont {S.}~\bibnamefont
  {De~Nicola}}, \bibinfo {author} {\bibfnamefont {L.}~\bibnamefont {Labate}},
  \bibinfo {author} {\bibfnamefont {P.}~\bibnamefont {Londrillo}}, \bibinfo
  {author} {\bibfnamefont {R.}~\bibnamefont {Fedele}}, \bibinfo {author}
  {\bibfnamefont {D.}~\bibnamefont {Terzani}}, \ and\ \bibinfo {author}
  {\bibfnamefont {L.~A.}\ \bibnamefont {Gizzi}},\ }\href@noop {} {\bibfield
  {journal} {\bibinfo  {journal} {Physics of Plasmas}\ }\textbf {\bibinfo
  {volume} {24}},\ \bibinfo {pages} {103120} (\bibinfo {year}
  {2017})}\BibitemShut {NoStop}%
\bibitem [{\citenamefont {Bourgeois}, \citenamefont {Cowley},\ and\
  \citenamefont {Hooker}(2013)}]{bourgeois2013two}%
  \BibitemOpen
  \bibfield  {author} {\bibinfo {author} {\bibfnamefont {N.}~\bibnamefont
  {Bourgeois}}, \bibinfo {author} {\bibfnamefont {J.}~\bibnamefont {Cowley}}, \
  and\ \bibinfo {author} {\bibfnamefont {S.~M.}\ \bibnamefont {Hooker}},\
  }\href@noop {} {\bibfield  {journal} {\bibinfo  {journal} {Physical Review
  Letters}\ }\textbf {\bibinfo {volume} {111}},\ \bibinfo {pages} {155004}
  (\bibinfo {year} {2013})}\BibitemShut {NoStop}%
\bibitem [{\citenamefont {Yu}\ \emph {et~al.}(2014)\citenamefont {Yu},
  \citenamefont {Esarey}, \citenamefont {Schroeder}, \citenamefont {Vay},
  \citenamefont {Benedetti}, \citenamefont {Geddes}, \citenamefont {Chen},\
  and\ \citenamefont {Leemans}}]{yu2014two}%
  \BibitemOpen
  \bibfield  {author} {\bibinfo {author} {\bibfnamefont {L.-L.}\ \bibnamefont
  {Yu}}, \bibinfo {author} {\bibfnamefont {E.}~\bibnamefont {Esarey}}, \bibinfo
  {author} {\bibfnamefont {C.}~\bibnamefont {Schroeder}}, \bibinfo {author}
  {\bibfnamefont {J.-L.}\ \bibnamefont {Vay}}, \bibinfo {author} {\bibfnamefont
  {C.}~\bibnamefont {Benedetti}}, \bibinfo {author} {\bibfnamefont
  {C.}~\bibnamefont {Geddes}}, \bibinfo {author} {\bibfnamefont
  {M.}~\bibnamefont {Chen}}, \ and\ \bibinfo {author} {\bibfnamefont
  {W.}~\bibnamefont {Leemans}},\ }\href@noop {} {\bibfield  {journal} {\bibinfo
   {journal} {Physical Review Letters}\ }\textbf {\bibinfo {volume} {112}},\
  \bibinfo {pages} {125001} (\bibinfo {year} {2014})}\BibitemShut {NoStop}%
\bibitem [{\citenamefont {Xu}\ \emph {et~al.}(2014)\citenamefont {Xu},
  \citenamefont {Wu}, \citenamefont {Zhang}, \citenamefont {Li}, \citenamefont
  {Wan}, \citenamefont {Hua}, \citenamefont {Pai}, \citenamefont {Lu},
  \citenamefont {Yu}, \citenamefont {Joshi} \emph {et~al.}}]{xu2014low}%
  \BibitemOpen
  \bibfield  {author} {\bibinfo {author} {\bibfnamefont {X.}~\bibnamefont
  {Xu}}, \bibinfo {author} {\bibfnamefont {Y.}~\bibnamefont {Wu}}, \bibinfo
  {author} {\bibfnamefont {C.}~\bibnamefont {Zhang}}, \bibinfo {author}
  {\bibfnamefont {F.}~\bibnamefont {Li}}, \bibinfo {author} {\bibfnamefont
  {Y.}~\bibnamefont {Wan}}, \bibinfo {author} {\bibfnamefont {J.}~\bibnamefont
  {Hua}}, \bibinfo {author} {\bibfnamefont {C.-H.}\ \bibnamefont {Pai}},
  \bibinfo {author} {\bibfnamefont {W.}~\bibnamefont {Lu}}, \bibinfo {author}
  {\bibfnamefont {P.}~\bibnamefont {Yu}}, \bibinfo {author} {\bibfnamefont
  {C.}~\bibnamefont {Joshi}},  \emph {et~al.},\ }\href@noop {} {\bibfield
  {journal} {\bibinfo  {journal} {Physical Review Special Topics-Accelerators
  and Beams}\ }\textbf {\bibinfo {volume} {17}},\ \bibinfo {pages} {061301}
  (\bibinfo {year} {2014})}\BibitemShut {NoStop}%
\bibitem [{\citenamefont {Chen}\ \emph {et~al.}(2013)\citenamefont {Chen},
  \citenamefont {Yan}, \citenamefont {Li}, \citenamefont {Hu}, \citenamefont
  {Zhang}, \citenamefont {Wang}, \citenamefont {Hafz}, \citenamefont {Mao},
  \citenamefont {Huang}, \citenamefont {Ma} \emph {et~al.}}]{chen2013bright}%
  \BibitemOpen
  \bibfield  {author} {\bibinfo {author} {\bibfnamefont {L.}~\bibnamefont
  {Chen}}, \bibinfo {author} {\bibfnamefont {W.}~\bibnamefont {Yan}}, \bibinfo
  {author} {\bibfnamefont {D.}~\bibnamefont {Li}}, \bibinfo {author}
  {\bibfnamefont {Z.}~\bibnamefont {Hu}}, \bibinfo {author} {\bibfnamefont
  {L.}~\bibnamefont {Zhang}}, \bibinfo {author} {\bibfnamefont
  {W.}~\bibnamefont {Wang}}, \bibinfo {author} {\bibfnamefont {N.}~\bibnamefont
  {Hafz}}, \bibinfo {author} {\bibfnamefont {J.}~\bibnamefont {Mao}}, \bibinfo
  {author} {\bibfnamefont {K.}~\bibnamefont {Huang}}, \bibinfo {author}
  {\bibfnamefont {Y.}~\bibnamefont {Ma}},  \emph {et~al.},\ }\href@noop {}
  {\bibfield  {journal} {\bibinfo  {journal} {Scientific reports}\ }\textbf
  {\bibinfo {volume} {3}},\ \bibinfo {pages} {1912} (\bibinfo {year}
  {2013})}\BibitemShut {NoStop}%
\bibitem [{\citenamefont {Gu{\'{e}}not}\ \emph {et~al.}(2017)\citenamefont
  {Gu{\'{e}}not}, \citenamefont {Gustas}, \citenamefont {Vernier},
  \citenamefont {Beaurepaire}, \citenamefont {B{\"{o}}hle}, \citenamefont
  {Bocoum}, \citenamefont {Lozano}, \citenamefont {Jullien}, \citenamefont
  {Lopez-Martens}, \citenamefont {Lifschitz},\ and\ \citenamefont
  {Faure}}]{guenot_natphot2017}%
  \BibitemOpen
  \bibfield  {author} {\bibinfo {author} {\bibfnamefont {D.}~\bibnamefont
  {Gu{\'{e}}not}}, \bibinfo {author} {\bibfnamefont {D.}~\bibnamefont
  {Gustas}}, \bibinfo {author} {\bibfnamefont {A.}~\bibnamefont {Vernier}},
  \bibinfo {author} {\bibfnamefont {B.}~\bibnamefont {Beaurepaire}}, \bibinfo
  {author} {\bibfnamefont {F.}~\bibnamefont {B{\"{o}}hle}}, \bibinfo {author}
  {\bibfnamefont {M.}~\bibnamefont {Bocoum}}, \bibinfo {author} {\bibfnamefont
  {M.}~\bibnamefont {Lozano}}, \bibinfo {author} {\bibfnamefont
  {A.}~\bibnamefont {Jullien}}, \bibinfo {author} {\bibfnamefont
  {R.}~\bibnamefont {Lopez-Martens}}, \bibinfo {author} {\bibfnamefont
  {A.}~\bibnamefont {Lifschitz}}, \ and\ \bibinfo {author} {\bibfnamefont
  {J.}~\bibnamefont {Faure}},\ }\bibfield  {title} {\enquote {\bibinfo {title}
  {{Relativistic electron beams driven by kHz single-cycle light pulses}},}\
  }\href {\doibase 10.1038/nphoton.2017.46} {\bibfield  {journal} {\bibinfo
  {journal} {Nature Photonics}\ }\textbf {\bibinfo {volume} {11}},\ \bibinfo
  {pages} {293--296} (\bibinfo {year} {2017})},\ \Eprint
  {http://arxiv.org/abs/1611.09543} {1611.09543} \BibitemShut {NoStop}%
\bibitem [{\citenamefont {{Forschungszentrum J\"ulich GmbH
  (PGI-6)}}(2019)}]{JUSPARC_fzj2019}%
  \BibitemOpen
  \bibfield  {author} {\bibinfo {author} {\bibnamefont {{Forschungszentrum
  J\"ulich GmbH (PGI-6)}}},\ }\bibfield  {title} {\enquote {\bibinfo {title}
  {Jusparc: Jülich short-pulsed particle acceleration and radiation centre},}\
  }\href@noop {} {\bibfield  {journal} {\bibinfo  {journal} {Journal of
  large-scale research facilities}\ } (\bibinfo {year} {submitted,
  2019})}\BibitemShut {NoStop}%
\bibitem [{\citenamefont {Rosenzweig}\ \emph {et~al.}(1991)\citenamefont
  {Rosenzweig}, \citenamefont {Breizman}, \citenamefont {Katsouleas},\ and\
  \citenamefont {Su}}]{rosenzweig_pra91}%
  \BibitemOpen
  \bibfield  {author} {\bibinfo {author} {\bibfnamefont {J.~B.}\ \bibnamefont
  {Rosenzweig}}, \bibinfo {author} {\bibfnamefont {B.}~\bibnamefont
  {Breizman}}, \bibinfo {author} {\bibfnamefont {T.}~\bibnamefont
  {Katsouleas}}, \ and\ \bibinfo {author} {\bibfnamefont {J.~J.}\ \bibnamefont
  {Su}},\ }\bibfield  {title} {\enquote {\bibinfo {title} {{Acceleration and
  focusing of electrons in two-dimensional nonlinear plasma wake fields}},}\
  }\href {\doibase 10.1103/PhysRevA.44.R6189} {\bibfield  {journal} {\bibinfo
  {journal} {Physical Review A}\ }\textbf {\bibinfo {volume} {44}},\ \bibinfo
  {pages} {R6189--R6192} (\bibinfo {year} {1991})}\BibitemShut {NoStop}%
\bibitem [{\citenamefont {Pukhov}\ and\ \citenamefont {Meyer-ter
  Vehn}(2002)}]{pukhov2002laser}%
  \BibitemOpen
  \bibfield  {author} {\bibinfo {author} {\bibfnamefont {A.}~\bibnamefont
  {Pukhov}}\ and\ \bibinfo {author} {\bibfnamefont {J.}~\bibnamefont {Meyer-ter
  Vehn}},\ }\href@noop {} {\bibfield  {journal} {\bibinfo  {journal} {Applied
  Physics B}\ }\textbf {\bibinfo {volume} {74}},\ \bibinfo {pages} {355--361}
  (\bibinfo {year} {2002})}\BibitemShut {NoStop}%
\bibitem [{\citenamefont {Kostyukov}, \citenamefont {Pukhov},\ and\
  \citenamefont {Kiselev}(2004)}]{kostyukov2004}%
  \BibitemOpen
  \bibfield  {author} {\bibinfo {author} {\bibfnamefont {I.}~\bibnamefont
  {Kostyukov}}, \bibinfo {author} {\bibfnamefont {A.}~\bibnamefont {Pukhov}}, \
  and\ \bibinfo {author} {\bibfnamefont {S.}~\bibnamefont {Kiselev}},\
  }\bibfield  {title} {\enquote {\bibinfo {title} {{Phenomenological theory of
  laser-plasma interaction in "bubble" regime}},}\ }\href {\doibase
  10.1063/1.1799371} {\bibfield  {journal} {\bibinfo  {journal} {Physics of
  Plasmas}\ }\textbf {\bibinfo {volume} {11}},\ \bibinfo {pages} {5256--5264}
  (\bibinfo {year} {2004})}\BibitemShut {NoStop}%
\bibitem [{\citenamefont {Lu}\ \emph {et~al.}(2006)\citenamefont {Lu},
  \citenamefont {Huang}, \citenamefont {Zhou}, \citenamefont {Mori},\ and\
  \citenamefont {Katsouleas}}]{lu2006nonlinear}%
  \BibitemOpen
  \bibfield  {author} {\bibinfo {author} {\bibfnamefont {W.}~\bibnamefont
  {Lu}}, \bibinfo {author} {\bibfnamefont {C.}~\bibnamefont {Huang}}, \bibinfo
  {author} {\bibfnamefont {M.}~\bibnamefont {Zhou}}, \bibinfo {author}
  {\bibfnamefont {W.~B.}\ \bibnamefont {Mori}}, \ and\ \bibinfo {author}
  {\bibfnamefont {T.}~\bibnamefont {Katsouleas}},\ }\bibfield  {title}
  {\enquote {\bibinfo {title} {{Nonlinear Theory for Relativistic Plasma
  Wakefields in the Blowout Regime}},}\ }\href {\doibase
  10.1103/PhysRevLett.96.165002} {\bibfield  {journal} {\bibinfo  {journal}
  {Physical Review Letters}\ }\textbf {\bibinfo {volume} {96}},\ \bibinfo
  {pages} {165002} (\bibinfo {year} {2006})}\BibitemShut {NoStop}%
\bibitem [{\citenamefont {Lu}\ \emph {et~al.}(2007)\citenamefont {Lu},
  \citenamefont {Tzoufras}, \citenamefont {Joshi}, \citenamefont {Tsung},
  \citenamefont {Mori}, \citenamefont {Vieira}, \citenamefont {Fonseca},\ and\
  \citenamefont {Silva}}]{lu2007generating}%
  \BibitemOpen
  \bibfield  {author} {\bibinfo {author} {\bibfnamefont {W.}~\bibnamefont
  {Lu}}, \bibinfo {author} {\bibfnamefont {M.}~\bibnamefont {Tzoufras}},
  \bibinfo {author} {\bibfnamefont {C.}~\bibnamefont {Joshi}}, \bibinfo
  {author} {\bibfnamefont {F.}~\bibnamefont {Tsung}}, \bibinfo {author}
  {\bibfnamefont {W.}~\bibnamefont {Mori}}, \bibinfo {author} {\bibfnamefont
  {J.}~\bibnamefont {Vieira}}, \bibinfo {author} {\bibfnamefont
  {R.}~\bibnamefont {Fonseca}}, \ and\ \bibinfo {author} {\bibfnamefont
  {L.}~\bibnamefont {Silva}},\ }\href@noop {} {\bibfield  {journal} {\bibinfo
  {journal} {Physical Review Special Topics-Accelerators and Beams}\ }\textbf
  {\bibinfo {volume} {10}},\ \bibinfo {pages} {061301} (\bibinfo {year}
  {2007})}\BibitemShut {NoStop}%
\bibitem [{\citenamefont {Esirkepov}\ \emph {et~al.}(2006)\citenamefont
  {Esirkepov}, \citenamefont {Bulanov}, \citenamefont {Yamagiwa},\ and\
  \citenamefont {Tajima}}]{esirkepov2006electron}%
  \BibitemOpen
  \bibfield  {author} {\bibinfo {author} {\bibfnamefont {T.}~\bibnamefont
  {Esirkepov}}, \bibinfo {author} {\bibfnamefont {S.}~\bibnamefont {Bulanov}},
  \bibinfo {author} {\bibfnamefont {M.}~\bibnamefont {Yamagiwa}}, \ and\
  \bibinfo {author} {\bibfnamefont {T.}~\bibnamefont {Tajima}},\ }\href@noop {}
  {\bibfield  {journal} {\bibinfo  {journal} {Physical Review Letters}\
  }\textbf {\bibinfo {volume} {96}},\ \bibinfo {pages} {014803} (\bibinfo
  {year} {2006})}\BibitemShut {NoStop}%
\bibitem [{\citenamefont {Corde}\ \emph {et~al.}(2013)\citenamefont {Corde},
  \citenamefont {Phuoc}, \citenamefont {Lambert}, \citenamefont {Fitour},
  \citenamefont {Malka}, \citenamefont {Rousse}, \citenamefont {Beck},\ and\
  \citenamefont {Lefebvre}}]{corde2013femtosecond}%
  \BibitemOpen
  \bibfield  {author} {\bibinfo {author} {\bibfnamefont {S.}~\bibnamefont
  {Corde}}, \bibinfo {author} {\bibfnamefont {K.~T.}\ \bibnamefont {Phuoc}},
  \bibinfo {author} {\bibfnamefont {G.}~\bibnamefont {Lambert}}, \bibinfo
  {author} {\bibfnamefont {R.}~\bibnamefont {Fitour}}, \bibinfo {author}
  {\bibfnamefont {V.}~\bibnamefont {Malka}}, \bibinfo {author} {\bibfnamefont
  {A.}~\bibnamefont {Rousse}}, \bibinfo {author} {\bibfnamefont
  {A.}~\bibnamefont {Beck}}, \ and\ \bibinfo {author} {\bibfnamefont
  {E.}~\bibnamefont {Lefebvre}},\ }\href@noop {} {\bibfield  {journal}
  {\bibinfo  {journal} {Reviews of Modern Physics}\ }\textbf {\bibinfo {volume}
  {85}},\ \bibinfo {pages} {1} (\bibinfo {year} {2013})}\BibitemShut {NoStop}%
\bibitem [{\citenamefont {Esarey}\ and\ \citenamefont
  {Pilloff}(1995)}]{esarey_pop95}%
  \BibitemOpen
  \bibfield  {author} {\bibinfo {author} {\bibfnamefont {E.}~\bibnamefont
  {Esarey}}\ and\ \bibinfo {author} {\bibfnamefont {M.}~\bibnamefont
  {Pilloff}},\ }\bibfield  {title} {\enquote {\bibinfo {title} {{Trapping and
  acceleration in nonlinear plasma waves}},}\ }\href {\doibase
  10.1063/1.871358} {\bibfield  {journal} {\bibinfo  {journal} {Physics of
  Plasmas}\ }\textbf {\bibinfo {volume} {2}},\ \bibinfo {pages} {1432--1436}
  (\bibinfo {year} {1995})}\BibitemShut {NoStop}%
\bibitem [{\citenamefont {Faure}(2017)}]{CERNYR218}%
  \BibitemOpen
  \bibfield  {author} {\bibinfo {author} {\bibfnamefont {J.}~\bibnamefont
  {Faure}},\ }\bibfield  {title} {\enquote {\bibinfo {title} {Plasma injection
  schemes for laser-plasma accelerators},}\ }\href@noop {} {\bibfield
  {journal} {\bibinfo  {journal} {CERN Yellow Reports}\ }\textbf {\bibinfo
  {volume} {1}},\ \bibinfo {pages} {143} (\bibinfo {year} {2017})}\BibitemShut
  {NoStop}%
\bibitem [{\citenamefont {Tsung}\ \emph {et~al.}(2006)\citenamefont {Tsung},
  \citenamefont {Lu}, \citenamefont {Tzoufras}, \citenamefont {Mori},
  \citenamefont {Joshi}, \citenamefont {Vieira}, \citenamefont {Silva},\ and\
  \citenamefont {Fonseca}}]{tsung2006simulation}%
  \BibitemOpen
  \bibfield  {author} {\bibinfo {author} {\bibfnamefont {F.}~\bibnamefont
  {Tsung}}, \bibinfo {author} {\bibfnamefont {W.}~\bibnamefont {Lu}}, \bibinfo
  {author} {\bibfnamefont {M.}~\bibnamefont {Tzoufras}}, \bibinfo {author}
  {\bibfnamefont {W.}~\bibnamefont {Mori}}, \bibinfo {author} {\bibfnamefont
  {C.}~\bibnamefont {Joshi}}, \bibinfo {author} {\bibfnamefont
  {J.}~\bibnamefont {Vieira}}, \bibinfo {author} {\bibfnamefont
  {L.}~\bibnamefont {Silva}}, \ and\ \bibinfo {author} {\bibfnamefont
  {R.}~\bibnamefont {Fonseca}},\ }\href@noop {} {\bibfield  {journal} {\bibinfo
   {journal} {Physics of Plasmas}\ }\textbf {\bibinfo {volume} {13}},\ \bibinfo
  {pages} {056708} (\bibinfo {year} {2006})}\BibitemShut {NoStop}%
\bibitem [{\citenamefont {Froula}\ \emph {et~al.}(2009)\citenamefont {Froula},
  \citenamefont {Clayton}, \citenamefont {D{\"o}ppner}, \citenamefont {Marsh},
  \citenamefont {Barty}, \citenamefont {Divol}, \citenamefont {Fonseca},
  \citenamefont {Glenzer}, \citenamefont {Joshi}, \citenamefont {Lu} \emph
  {et~al.}}]{froula2009measurements}%
  \BibitemOpen
  \bibfield  {author} {\bibinfo {author} {\bibfnamefont {D.}~\bibnamefont
  {Froula}}, \bibinfo {author} {\bibfnamefont {C.}~\bibnamefont {Clayton}},
  \bibinfo {author} {\bibfnamefont {T.}~\bibnamefont {D{\"o}ppner}}, \bibinfo
  {author} {\bibfnamefont {K.}~\bibnamefont {Marsh}}, \bibinfo {author}
  {\bibfnamefont {C.}~\bibnamefont {Barty}}, \bibinfo {author} {\bibfnamefont
  {L.}~\bibnamefont {Divol}}, \bibinfo {author} {\bibfnamefont
  {R.}~\bibnamefont {Fonseca}}, \bibinfo {author} {\bibfnamefont
  {S.}~\bibnamefont {Glenzer}}, \bibinfo {author} {\bibfnamefont
  {C.}~\bibnamefont {Joshi}}, \bibinfo {author} {\bibfnamefont
  {W.}~\bibnamefont {Lu}},  \emph {et~al.},\ }\bibfield  {title} {\enquote
  {\bibinfo {title} {Measurements of the critical power for self-injection of
  electrons in a laser wakefield accelerator},}\ }\href@noop {} {\bibfield
  {journal} {\bibinfo  {journal} {Physical review letters}\ }\textbf {\bibinfo
  {volume} {103}},\ \bibinfo {pages} {215006} (\bibinfo {year}
  {2009})}\BibitemShut {NoStop}%
\bibitem [{\citenamefont {Mangles}\ \emph {et~al.}(2012)\citenamefont
  {Mangles}, \citenamefont {Genoud}, \citenamefont {Bloom}, \citenamefont
  {Burza}, \citenamefont {Najmudin}, \citenamefont {Persson}, \citenamefont
  {Svensson}, \citenamefont {Thomas},\ and\ \citenamefont
  {Wahlstr{\"o}m}}]{mangles2012self}%
  \BibitemOpen
  \bibfield  {author} {\bibinfo {author} {\bibfnamefont {S.~P.}\ \bibnamefont
  {Mangles}}, \bibinfo {author} {\bibfnamefont {G.}~\bibnamefont {Genoud}},
  \bibinfo {author} {\bibfnamefont {M.~S.}\ \bibnamefont {Bloom}}, \bibinfo
  {author} {\bibfnamefont {M.}~\bibnamefont {Burza}}, \bibinfo {author}
  {\bibfnamefont {Z.}~\bibnamefont {Najmudin}}, \bibinfo {author}
  {\bibfnamefont {A.}~\bibnamefont {Persson}}, \bibinfo {author} {\bibfnamefont
  {K.}~\bibnamefont {Svensson}}, \bibinfo {author} {\bibfnamefont {A.~G.}\
  \bibnamefont {Thomas}}, \ and\ \bibinfo {author} {\bibfnamefont {C.-G.}\
  \bibnamefont {Wahlstr{\"o}m}},\ }\bibfield  {title} {\enquote {\bibinfo
  {title} {Self-injection threshold in self-guided laser wakefield
  accelerators},}\ }\href@noop {} {\bibfield  {journal} {\bibinfo  {journal}
  {Physical Review Special Topics-Accelerators and Beams}\ }\textbf {\bibinfo
  {volume} {15}},\ \bibinfo {pages} {011302} (\bibinfo {year}
  {2012})}\BibitemShut {NoStop}%
\bibitem [{\citenamefont {Pak}\ \emph {et~al.}(2010)\citenamefont {Pak},
  \citenamefont {Marsh}, \citenamefont {Martins}, \citenamefont {Lu},
  \citenamefont {Mori},\ and\ \citenamefont {Joshi}}]{pak2010injection}%
  \BibitemOpen
  \bibfield  {author} {\bibinfo {author} {\bibfnamefont {A.}~\bibnamefont
  {Pak}}, \bibinfo {author} {\bibfnamefont {K.}~\bibnamefont {Marsh}}, \bibinfo
  {author} {\bibfnamefont {S.}~\bibnamefont {Martins}}, \bibinfo {author}
  {\bibfnamefont {W.}~\bibnamefont {Lu}}, \bibinfo {author} {\bibfnamefont
  {W.}~\bibnamefont {Mori}}, \ and\ \bibinfo {author} {\bibfnamefont
  {C.}~\bibnamefont {Joshi}},\ }\href@noop {} {\bibfield  {journal} {\bibinfo
  {journal} {Physical Review Letters}\ }\textbf {\bibinfo {volume} {104}},\
  \bibinfo {pages} {025003} (\bibinfo {year} {2010})}\BibitemShut {NoStop}%
\bibitem [{\citenamefont {McGuffey}\ \emph {et~al.}(2010)\citenamefont
  {McGuffey}, \citenamefont {Thomas}, \citenamefont {Schumaker}, \citenamefont
  {Matsuoka}, \citenamefont {Chvykov}, \citenamefont {Dollar}, \citenamefont
  {Kalintchenko}, \citenamefont {Yanovsky}, \citenamefont {Maksimchuk},
  \citenamefont {Krushelnick} \emph {et~al.}}]{mcguffey2010ionization}%
  \BibitemOpen
  \bibfield  {author} {\bibinfo {author} {\bibfnamefont {C.}~\bibnamefont
  {McGuffey}}, \bibinfo {author} {\bibfnamefont {A.}~\bibnamefont {Thomas}},
  \bibinfo {author} {\bibfnamefont {W.}~\bibnamefont {Schumaker}}, \bibinfo
  {author} {\bibfnamefont {T.}~\bibnamefont {Matsuoka}}, \bibinfo {author}
  {\bibfnamefont {V.}~\bibnamefont {Chvykov}}, \bibinfo {author} {\bibfnamefont
  {F.}~\bibnamefont {Dollar}}, \bibinfo {author} {\bibfnamefont
  {G.}~\bibnamefont {Kalintchenko}}, \bibinfo {author} {\bibfnamefont
  {V.}~\bibnamefont {Yanovsky}}, \bibinfo {author} {\bibfnamefont
  {A.}~\bibnamefont {Maksimchuk}}, \bibinfo {author} {\bibfnamefont
  {K.}~\bibnamefont {Krushelnick}},  \emph {et~al.},\ }\href@noop {} {\bibfield
   {journal} {\bibinfo  {journal} {Physical Review Letters}\ }\textbf {\bibinfo
  {volume} {104}},\ \bibinfo {pages} {025004} (\bibinfo {year}
  {2010})}\BibitemShut {NoStop}%
\bibitem [{\citenamefont {Clayton}\ \emph {et~al.}(2010)\citenamefont
  {Clayton}, \citenamefont {Ralph}, \citenamefont {Albert}, \citenamefont
  {Fonseca}, \citenamefont {Glenzer}, \citenamefont {Joshi}, \citenamefont
  {Lu}, \citenamefont {Marsh}, \citenamefont {Martins}, \citenamefont {Mori}
  \emph {et~al.}}]{clayton2010self}%
  \BibitemOpen
  \bibfield  {author} {\bibinfo {author} {\bibfnamefont {C.}~\bibnamefont
  {Clayton}}, \bibinfo {author} {\bibfnamefont {J.}~\bibnamefont {Ralph}},
  \bibinfo {author} {\bibfnamefont {F.}~\bibnamefont {Albert}}, \bibinfo
  {author} {\bibfnamefont {R.}~\bibnamefont {Fonseca}}, \bibinfo {author}
  {\bibfnamefont {S.}~\bibnamefont {Glenzer}}, \bibinfo {author} {\bibfnamefont
  {C.}~\bibnamefont {Joshi}}, \bibinfo {author} {\bibfnamefont
  {W.}~\bibnamefont {Lu}}, \bibinfo {author} {\bibfnamefont {K.}~\bibnamefont
  {Marsh}}, \bibinfo {author} {\bibfnamefont {S.}~\bibnamefont {Martins}},
  \bibinfo {author} {\bibfnamefont {W.}~\bibnamefont {Mori}},  \emph {et~al.},\
  }\href@noop {} {\bibfield  {journal} {\bibinfo  {journal} {Physical Review
  Letters}\ }\textbf {\bibinfo {volume} {105}},\ \bibinfo {pages} {105003}
  (\bibinfo {year} {2010})}\BibitemShut {NoStop}%
\bibitem [{\citenamefont {Faure}\ \emph {et~al.}(2006)\citenamefont {Faure},
  \citenamefont {Rechatin}, \citenamefont {Norlin}, \citenamefont {Lifschitz},
  \citenamefont {Glinec},\ and\ \citenamefont {Malka}}]{faure2006controlled}%
  \BibitemOpen
  \bibfield  {author} {\bibinfo {author} {\bibfnamefont {J.}~\bibnamefont
  {Faure}}, \bibinfo {author} {\bibfnamefont {C.}~\bibnamefont {Rechatin}},
  \bibinfo {author} {\bibfnamefont {A.}~\bibnamefont {Norlin}}, \bibinfo
  {author} {\bibfnamefont {A.}~\bibnamefont {Lifschitz}}, \bibinfo {author}
  {\bibfnamefont {Y.}~\bibnamefont {Glinec}}, \ and\ \bibinfo {author}
  {\bibfnamefont {V.}~\bibnamefont {Malka}},\ }\href@noop {} {\bibfield
  {journal} {\bibinfo  {journal} {Nature}\ }\textbf {\bibinfo {volume} {444}},\
  \bibinfo {pages} {737} (\bibinfo {year} {2006})}\BibitemShut {NoStop}%
\bibitem [{\citenamefont {Rechatin}\ \emph {et~al.}(2009)\citenamefont
  {Rechatin}, \citenamefont {Faure}, \citenamefont {Ben-Isma{\"\i}l},
  \citenamefont {Lim}, \citenamefont {Fitour}, \citenamefont {Specka},
  \citenamefont {Videau}, \citenamefont {Tafzi}, \citenamefont {Burgy},\ and\
  \citenamefont {Malka}}]{rechatin2009controlling}%
  \BibitemOpen
  \bibfield  {author} {\bibinfo {author} {\bibfnamefont {C.}~\bibnamefont
  {Rechatin}}, \bibinfo {author} {\bibfnamefont {J.}~\bibnamefont {Faure}},
  \bibinfo {author} {\bibfnamefont {A.}~\bibnamefont {Ben-Isma{\"\i}l}},
  \bibinfo {author} {\bibfnamefont {J.}~\bibnamefont {Lim}}, \bibinfo {author}
  {\bibfnamefont {R.}~\bibnamefont {Fitour}}, \bibinfo {author} {\bibfnamefont
  {A.}~\bibnamefont {Specka}}, \bibinfo {author} {\bibfnamefont
  {H.}~\bibnamefont {Videau}}, \bibinfo {author} {\bibfnamefont
  {A.}~\bibnamefont {Tafzi}}, \bibinfo {author} {\bibfnamefont
  {F.}~\bibnamefont {Burgy}}, \ and\ \bibinfo {author} {\bibfnamefont
  {V.}~\bibnamefont {Malka}},\ }\href@noop {} {\bibfield  {journal} {\bibinfo
  {journal} {Physical Review Letters}\ }\textbf {\bibinfo {volume} {102}},\
  \bibinfo {pages} {164801} (\bibinfo {year} {2009})}\BibitemShut {NoStop}%
\bibitem [{\citenamefont {Lundh}\ \emph {et~al.}(2011)\citenamefont {Lundh},
  \citenamefont {Lim}, \citenamefont {Rechatin}, \citenamefont {Ammoura},
  \citenamefont {Ben-Isma{\"\i}l}, \citenamefont {Davoine}, \citenamefont
  {Gallot}, \citenamefont {Goddet}, \citenamefont {Lefebvre}, \citenamefont
  {Malka} \emph {et~al.}}]{lundh2011few}%
  \BibitemOpen
  \bibfield  {author} {\bibinfo {author} {\bibfnamefont {O.}~\bibnamefont
  {Lundh}}, \bibinfo {author} {\bibfnamefont {J.}~\bibnamefont {Lim}}, \bibinfo
  {author} {\bibfnamefont {C.}~\bibnamefont {Rechatin}}, \bibinfo {author}
  {\bibfnamefont {L.}~\bibnamefont {Ammoura}}, \bibinfo {author} {\bibfnamefont
  {A.}~\bibnamefont {Ben-Isma{\"\i}l}}, \bibinfo {author} {\bibfnamefont
  {X.}~\bibnamefont {Davoine}}, \bibinfo {author} {\bibfnamefont
  {G.}~\bibnamefont {Gallot}}, \bibinfo {author} {\bibfnamefont {J.-P.}\
  \bibnamefont {Goddet}}, \bibinfo {author} {\bibfnamefont {E.}~\bibnamefont
  {Lefebvre}}, \bibinfo {author} {\bibfnamefont {V.}~\bibnamefont {Malka}},
  \emph {et~al.},\ }\href@noop {} {\bibfield  {journal} {\bibinfo  {journal}
  {Nature Physics}\ }\textbf {\bibinfo {volume} {7}},\ \bibinfo {pages} {219}
  (\bibinfo {year} {2011})}\BibitemShut {NoStop}%
\bibitem [{\citenamefont {Brantov}\ \emph {et~al.}(2008)\citenamefont
  {Brantov}, \citenamefont {Esirkepov}, \citenamefont {Kando}, \citenamefont
  {Kotaki}, \citenamefont {Bychenkov},\ and\ \citenamefont
  {Bulanov}}]{brantov2008controlled}%
  \BibitemOpen
  \bibfield  {author} {\bibinfo {author} {\bibfnamefont {A.}~\bibnamefont
  {Brantov}}, \bibinfo {author} {\bibfnamefont {T.~Z.}\ \bibnamefont
  {Esirkepov}}, \bibinfo {author} {\bibfnamefont {M.}~\bibnamefont {Kando}},
  \bibinfo {author} {\bibfnamefont {H.}~\bibnamefont {Kotaki}}, \bibinfo
  {author} {\bibfnamefont {V.~Y.}\ \bibnamefont {Bychenkov}}, \ and\ \bibinfo
  {author} {\bibfnamefont {S.}~\bibnamefont {Bulanov}},\ }\href@noop {}
  {\bibfield  {journal} {\bibinfo  {journal} {Physics of Plasmas}\ }\textbf
  {\bibinfo {volume} {15}},\ \bibinfo {pages} {073111} (\bibinfo {year}
  {2008})}\BibitemShut {NoStop}%
\bibitem [{\citenamefont {Geddes}\ \emph {et~al.}(2008)\citenamefont {Geddes},
  \citenamefont {Nakamura}, \citenamefont {Plateau}, \citenamefont {Toth},
  \citenamefont {Cormier-Michel}, \citenamefont {Esarey}, \citenamefont
  {Schroeder}, \citenamefont {Cary},\ and\ \citenamefont
  {Leemans}}]{geddes2008plasma}%
  \BibitemOpen
  \bibfield  {author} {\bibinfo {author} {\bibfnamefont {C.}~\bibnamefont
  {Geddes}}, \bibinfo {author} {\bibfnamefont {K.}~\bibnamefont {Nakamura}},
  \bibinfo {author} {\bibfnamefont {G.}~\bibnamefont {Plateau}}, \bibinfo
  {author} {\bibfnamefont {C.}~\bibnamefont {Toth}}, \bibinfo {author}
  {\bibfnamefont {E.}~\bibnamefont {Cormier-Michel}}, \bibinfo {author}
  {\bibfnamefont {E.}~\bibnamefont {Esarey}}, \bibinfo {author} {\bibfnamefont
  {C.}~\bibnamefont {Schroeder}}, \bibinfo {author} {\bibfnamefont
  {J.}~\bibnamefont {Cary}}, \ and\ \bibinfo {author} {\bibfnamefont
  {W.}~\bibnamefont {Leemans}},\ }\href@noop {} {\bibfield  {journal} {\bibinfo
   {journal} {Physical Review Letters}\ }\textbf {\bibinfo {volume} {100}},\
  \bibinfo {pages} {215004} (\bibinfo {year} {2008})}\BibitemShut {NoStop}%
\bibitem [{\citenamefont {Schmid}\ \emph {et~al.}(2010)\citenamefont {Schmid},
  \citenamefont {Buck}, \citenamefont {Sears}, \citenamefont {Mikhailova},
  \citenamefont {Tautz}, \citenamefont {Herrmann}, \citenamefont {Geissler},
  \citenamefont {Krausz},\ and\ \citenamefont {Veisz}}]{schmid2010density}%
  \BibitemOpen
  \bibfield  {author} {\bibinfo {author} {\bibfnamefont {K.}~\bibnamefont
  {Schmid}}, \bibinfo {author} {\bibfnamefont {A.}~\bibnamefont {Buck}},
  \bibinfo {author} {\bibfnamefont {C.~M.}\ \bibnamefont {Sears}}, \bibinfo
  {author} {\bibfnamefont {J.~M.}\ \bibnamefont {Mikhailova}}, \bibinfo
  {author} {\bibfnamefont {R.}~\bibnamefont {Tautz}}, \bibinfo {author}
  {\bibfnamefont {D.}~\bibnamefont {Herrmann}}, \bibinfo {author}
  {\bibfnamefont {M.}~\bibnamefont {Geissler}}, \bibinfo {author}
  {\bibfnamefont {F.}~\bibnamefont {Krausz}}, \ and\ \bibinfo {author}
  {\bibfnamefont {L.}~\bibnamefont {Veisz}},\ }\href@noop {} {\bibfield
  {journal} {\bibinfo  {journal} {Physical Review Special Topics-Accelerators
  and Beams}\ }\textbf {\bibinfo {volume} {13}},\ \bibinfo {pages} {091301}
  (\bibinfo {year} {2010})}\BibitemShut {NoStop}%
\bibitem [{\citenamefont {He}\ \emph {et~al.}(2013)\citenamefont {He},
  \citenamefont {Hou}, \citenamefont {Nees}, \citenamefont {Easter},
  \citenamefont {Faure}, \citenamefont {Krushelnick},\ and\ \citenamefont
  {Thomas}}]{he2013high}%
  \BibitemOpen
  \bibfield  {author} {\bibinfo {author} {\bibfnamefont {Z.}~\bibnamefont
  {He}}, \bibinfo {author} {\bibfnamefont {B.}~\bibnamefont {Hou}}, \bibinfo
  {author} {\bibfnamefont {J.}~\bibnamefont {Nees}}, \bibinfo {author}
  {\bibfnamefont {J.}~\bibnamefont {Easter}}, \bibinfo {author} {\bibfnamefont
  {J.}~\bibnamefont {Faure}}, \bibinfo {author} {\bibfnamefont
  {K.}~\bibnamefont {Krushelnick}}, \ and\ \bibinfo {author} {\bibfnamefont
  {A.}~\bibnamefont {Thomas}},\ }\href@noop {} {\bibfield  {journal} {\bibinfo
  {journal} {New Journal of Physics}\ }\textbf {\bibinfo {volume} {15}},\
  \bibinfo {pages} {053016} (\bibinfo {year} {2013})}\BibitemShut {NoStop}%
\bibitem [{\citenamefont {Buck}\ \emph {et~al.}(2013)\citenamefont {Buck},
  \citenamefont {Wenz}, \citenamefont {Xu}, \citenamefont {Khrennikov},
  \citenamefont {Schmid}, \citenamefont {Heigoldt}, \citenamefont {Mikhailova},
  \citenamefont {Geissler}, \citenamefont {Shen}, \citenamefont {Krausz} \emph
  {et~al.}}]{buck2013shock}%
  \BibitemOpen
  \bibfield  {author} {\bibinfo {author} {\bibfnamefont {A.}~\bibnamefont
  {Buck}}, \bibinfo {author} {\bibfnamefont {J.}~\bibnamefont {Wenz}}, \bibinfo
  {author} {\bibfnamefont {J.}~\bibnamefont {Xu}}, \bibinfo {author}
  {\bibfnamefont {K.}~\bibnamefont {Khrennikov}}, \bibinfo {author}
  {\bibfnamefont {K.}~\bibnamefont {Schmid}}, \bibinfo {author} {\bibfnamefont
  {M.}~\bibnamefont {Heigoldt}}, \bibinfo {author} {\bibfnamefont {J.~M.}\
  \bibnamefont {Mikhailova}}, \bibinfo {author} {\bibfnamefont
  {M.}~\bibnamefont {Geissler}}, \bibinfo {author} {\bibfnamefont
  {B.}~\bibnamefont {Shen}}, \bibinfo {author} {\bibfnamefont {F.}~\bibnamefont
  {Krausz}},  \emph {et~al.},\ }\href@noop {} {\bibfield  {journal} {\bibinfo
  {journal} {Physical Review Letters}\ }\textbf {\bibinfo {volume} {110}},\
  \bibinfo {pages} {185006} (\bibinfo {year} {2013})}\BibitemShut {NoStop}%
\bibitem [{\citenamefont {Pathak}\ \emph {et~al.}(2018)\citenamefont {Pathak},
  \citenamefont {Kim}, \citenamefont {Vieira}, \citenamefont {Silva},\ and\
  \citenamefont {Nam}}]{pathak2018all}%
  \BibitemOpen
  \bibfield  {author} {\bibinfo {author} {\bibfnamefont {V.~B.}\ \bibnamefont
  {Pathak}}, \bibinfo {author} {\bibfnamefont {H.~T.}\ \bibnamefont {Kim}},
  \bibinfo {author} {\bibfnamefont {J.}~\bibnamefont {Vieira}}, \bibinfo
  {author} {\bibfnamefont {L.}~\bibnamefont {Silva}}, \ and\ \bibinfo {author}
  {\bibfnamefont {C.~H.}\ \bibnamefont {Nam}},\ }\bibfield  {title} {\enquote
  {\bibinfo {title} {All optical dual stage laser wakefield acceleration driven
  by two-color laser pulses},}\ }\href@noop {} {\bibfield  {journal} {\bibinfo
  {journal} {Scientific reports}\ }\textbf {\bibinfo {volume} {8}},\ \bibinfo
  {pages} {11772} (\bibinfo {year} {2018})}\BibitemShut {NoStop}%
\bibitem [{\citenamefont {Kalmykov}\ \emph {et~al.}(2018)\citenamefont
  {Kalmykov}, \citenamefont {Davoine}, \citenamefont {Ghebregziabher},\ and\
  \citenamefont {Shadwick}}]{Kalmykov_2018}%
  \BibitemOpen
  \bibfield  {author} {\bibinfo {author} {\bibfnamefont {S.~Y.}\ \bibnamefont
  {Kalmykov}}, \bibinfo {author} {\bibfnamefont {X.}~\bibnamefont {Davoine}},
  \bibinfo {author} {\bibfnamefont {I.}~\bibnamefont {Ghebregziabher}}, \ and\
  \bibinfo {author} {\bibfnamefont {B.~A.}\ \bibnamefont {Shadwick}},\
  }\bibfield  {title} {\enquote {\bibinfo {title} {Optically controlled
  laser{\textendash}plasma electron accelerator for compact gamma-ray
  sources},}\ }\href {\doibase 10.1088/1367-2630/aaad57} {\bibfield  {journal}
  {\bibinfo  {journal} {New Journal of Physics}\ }\textbf {\bibinfo {volume}
  {20}},\ \bibinfo {pages} {023047} (\bibinfo {year} {2018})}\BibitemShut
  {NoStop}%
\bibitem [{\citenamefont {Thomas}(2008)}]{thomas_prl2008}%
  \BibitemOpen
  \bibfield  {author} {\bibinfo {author} {\bibfnamefont {A.}~\bibnamefont
  {Thomas}},\ }\bibfield  {title} {\enquote {\bibinfo {title} {{Monoenergetic
  electronic beam production using dual collinear laser pulses}},}\ }\href
  {\doibase 10.1103/PhysRevLett.100.255002} {\bibfield  {journal} {\bibinfo
  {journal} {Physical Review Letters}\ }\textbf {\bibinfo {volume} {100}},\
  \bibinfo {pages} {16--19} (\bibinfo {year} {2008})}\BibitemShut {NoStop}%
\bibitem [{\citenamefont {Horn{\'{y}}}\ \emph {et~al.}(2018)\citenamefont
  {Horn{\'{y}}}, \citenamefont {Ma{\v{s}}l{\'{a}}rov{\'{a}}}, \citenamefont
  {Petr{\v{z}}{\'{i}}lka}, \citenamefont {Klimo}, \citenamefont
  {Kozlov{\'{a}}},\ and\ \citenamefont {Krůs}}]{horny_ppcf2018}%
  \BibitemOpen
  \bibfield  {author} {\bibinfo {author} {\bibfnamefont {V.}~\bibnamefont
  {Horn{\'{y}}}}, \bibinfo {author} {\bibfnamefont {D.}~\bibnamefont
  {Ma{\v{s}}l{\'{a}}rov{\'{a}}}}, \bibinfo {author} {\bibfnamefont
  {V.}~\bibnamefont {Petr{\v{z}}{\'{i}}lka}}, \bibinfo {author} {\bibfnamefont
  {O.}~\bibnamefont {Klimo}}, \bibinfo {author} {\bibfnamefont
  {M.}~\bibnamefont {Kozlov{\'{a}}}}, \ and\ \bibinfo {author} {\bibfnamefont
  {M.}~\bibnamefont {Krůs}},\ }\bibfield  {title} {\enquote {\bibinfo {title}
  {{Optical injection dynamics in two laser wakefield acceleration
  configurations}},}\ }\href {\doibase 10.1088/1361-6587/aabd07} {\bibfield
  {journal} {\bibinfo  {journal} {Plasma Physics and Controlled Fusion}\
  }\textbf {\bibinfo {volume} {60}},\ \bibinfo {pages} {064009} (\bibinfo
  {year} {2018})}\BibitemShut {NoStop}%
\bibitem [{\citenamefont {Fajardo}\ \emph {et~al.}(2017)\citenamefont
  {Fajardo}, \citenamefont {Westerhof}, \citenamefont {Riconda}, \citenamefont
  {Melzer}, \citenamefont {Bret},\ and\ \citenamefont {Dromey}}]{horny2017}%
  \BibitemOpen
  \bibinfo {editor} {\bibfnamefont {M.}~\bibnamefont {Fajardo}}, \bibinfo
  {editor} {\bibfnamefont {E.}~\bibnamefont {Westerhof}}, \bibinfo {editor}
  {\bibfnamefont {C.}~\bibnamefont {Riconda}}, \bibinfo {editor} {\bibfnamefont
  {A.}~\bibnamefont {Melzer}}, \bibinfo {editor} {\bibfnamefont
  {A.}~\bibnamefont {Bret}}, \ and\ \bibinfo {editor} {\bibfnamefont
  {B.}~\bibnamefont {Dromey}},\ eds.,\ \href {arXiv preprint arXiv:1807.07845
  (2018)} {\emph {\bibinfo {title} {Proc. 44th EPS Conference on Plasma
  Physics}}},\ Vol.\ \bibinfo {volume} {41F}\ (\bibinfo  {publisher} {European
  Physical Society},\ \bibinfo {address} {Belfast},\ \bibinfo {year}
  {2017})\BibitemShut {NoStop}%
\bibitem [{\citenamefont {Chitgar}\ \emph {et~al.}(2018)\citenamefont
  {Chitgar}, \citenamefont {Gibbon}, \citenamefont {B\"oker}, \citenamefont
  {Lehrach},\ and\ \citenamefont {Büscher}}]{chitgar2018}%
  \BibitemOpen
  \bibfield  {author} {\bibinfo {author} {\bibfnamefont {Z.~M.}\ \bibnamefont
  {Chitgar}}, \bibinfo {author} {\bibfnamefont {P.}~\bibnamefont {Gibbon}},
  \bibinfo {author} {\bibfnamefont {J.}~\bibnamefont {B\"oker}}, \bibinfo
  {author} {\bibfnamefont {A.}~\bibnamefont {Lehrach}}, \ and\ \bibinfo
  {author} {\bibfnamefont {M.}~\bibnamefont {Büscher}},\ }\bibfield  {title}
  {\enquote {\bibinfo {title} {Enhanced betatron-radiation energy using two
  collinear laser pulses},}\ }in\ \href {arXiv preprint arXiv:1807.07845 (2018)
  % http://ocs.ciemat.es/EPS2018PAP/pdf/P2.2006.pdf} {\emph {\bibinfo
  {booktitle} {45th EPS Conference on Plasma Physics}}},\ Vol.\ \bibinfo
  {volume} {42A},\ \bibinfo {editor} {edited by\ \bibinfo {editor}
  {\bibfnamefont {S.}~\bibnamefont {Coda}}, \bibinfo {editor} {\bibfnamefont
  {J.}~\bibnamefont {Berndt}}, \bibinfo {editor} {\bibfnamefont
  {G.}~\bibnamefont {Lapenta}}, \bibinfo {editor} {\bibfnamefont
  {M.}~\bibnamefont {Mantsinen}}, \bibinfo {editor} {\bibfnamefont
  {C.}~\bibnamefont {Michaut}}, \ and\ \bibinfo {editor} {\bibfnamefont
  {S.}~\bibnamefont {Weber}}}\ (\bibinfo  {publisher} {European Physical
  Society},\ \bibinfo {address} {Prague},\ \bibinfo {year} {2018})\ p.\
  \bibinfo {pages} {P2.2006}\BibitemShut {NoStop}%
\bibitem [{\citenamefont {Arber}\ \emph {et~al.}(2015)\citenamefont {Arber},
  \citenamefont {Bennett}, \citenamefont {Brady}, \citenamefont
  {Lawrence-Douglas}, \citenamefont {Ramsay}, \citenamefont {Sircombe},
  \citenamefont {Gillies}, \citenamefont {Evans}, \citenamefont {Schmitz},
  \citenamefont {Bell} \emph {et~al.}}]{arber2015contemporary}%
  \BibitemOpen
  \bibfield  {author} {\bibinfo {author} {\bibfnamefont {T.}~\bibnamefont
  {Arber}}, \bibinfo {author} {\bibfnamefont {K.}~\bibnamefont {Bennett}},
  \bibinfo {author} {\bibfnamefont {C.}~\bibnamefont {Brady}}, \bibinfo
  {author} {\bibfnamefont {A.}~\bibnamefont {Lawrence-Douglas}}, \bibinfo
  {author} {\bibfnamefont {M.}~\bibnamefont {Ramsay}}, \bibinfo {author}
  {\bibfnamefont {N.}~\bibnamefont {Sircombe}}, \bibinfo {author}
  {\bibfnamefont {P.}~\bibnamefont {Gillies}}, \bibinfo {author} {\bibfnamefont
  {R.}~\bibnamefont {Evans}}, \bibinfo {author} {\bibfnamefont
  {H.}~\bibnamefont {Schmitz}}, \bibinfo {author} {\bibfnamefont
  {A.}~\bibnamefont {Bell}},  \emph {et~al.},\ }\bibfield  {title} {\enquote
  {\bibinfo {title} {Contemporary particle-in-cell approach to laser-plasma
  modelling},}\ }\href@noop {} {\bibfield  {journal} {\bibinfo  {journal}
  {Plasma Physics and Controlled Fusion}\ }\textbf {\bibinfo {volume} {57}},\
  \bibinfo {pages} {113001} (\bibinfo {year} {2015})}\BibitemShut {NoStop}%
\bibitem [{\citenamefont {Kalmykov}\ \emph {et~al.}(2009)\citenamefont
  {Kalmykov}, \citenamefont {Yi}, \citenamefont {Khudik},\ and\ \citenamefont
  {Shvets}}]{kalmykov2009electron}%
  \BibitemOpen
  \bibfield  {author} {\bibinfo {author} {\bibfnamefont {S.}~\bibnamefont
  {Kalmykov}}, \bibinfo {author} {\bibfnamefont {S.}~\bibnamefont {Yi}},
  \bibinfo {author} {\bibfnamefont {V.}~\bibnamefont {Khudik}}, \ and\ \bibinfo
  {author} {\bibfnamefont {G.}~\bibnamefont {Shvets}},\ }\bibfield  {title}
  {\enquote {\bibinfo {title} {Electron self-injection and trapping into an
  evolving plasma bubble},}\ }\href@noop {} {\bibfield  {journal} {\bibinfo
  {journal} {Physical Review Letters}\ }\textbf {\bibinfo {volume} {103}},\
  \bibinfo {pages} {135004} (\bibinfo {year} {2009})}\BibitemShut {NoStop}%
\bibitem [{\citenamefont {Benedetti}\ \emph {et~al.}(2013)\citenamefont
  {Benedetti}, \citenamefont {Schroeder}, \citenamefont {Esarey}, \citenamefont
  {Rossi},\ and\ \citenamefont {Leemans}}]{benedetti2013numerical}%
  \BibitemOpen
  \bibfield  {author} {\bibinfo {author} {\bibfnamefont {C.}~\bibnamefont
  {Benedetti}}, \bibinfo {author} {\bibfnamefont {C.}~\bibnamefont
  {Schroeder}}, \bibinfo {author} {\bibfnamefont {E.}~\bibnamefont {Esarey}},
  \bibinfo {author} {\bibfnamefont {F.}~\bibnamefont {Rossi}}, \ and\ \bibinfo
  {author} {\bibfnamefont {W.}~\bibnamefont {Leemans}},\ }\bibfield  {title}
  {\enquote {\bibinfo {title} {Numerical investigation of electron
  self-injection in the nonlinear bubble regime},}\ }\href@noop {} {\bibfield
  {journal} {\bibinfo  {journal} {Physics of Plasmas}\ }\textbf {\bibinfo
  {volume} {20}},\ \bibinfo {pages} {103108} (\bibinfo {year}
  {2013})}\BibitemShut {NoStop}%
\bibitem [{\citenamefont {Thomas}(2010)}]{thomas2010scalings}%
  \BibitemOpen
  \bibfield  {author} {\bibinfo {author} {\bibfnamefont {A.~G.~R.}\
  \bibnamefont {Thomas}},\ }\bibfield  {title} {\enquote {\bibinfo {title}
  {Scalings for radiation from plasma bubbles},}\ }\href@noop {} {\bibfield
  {journal} {\bibinfo  {journal} {Physics of Plasmas}\ }\textbf {\bibinfo
  {volume} {17}},\ \bibinfo {pages} {056708} (\bibinfo {year}
  {2010})}\BibitemShut {NoStop}%
\bibitem [{\citenamefont {Yi}\ \emph {et~al.}(2011)\citenamefont {Yi},
  \citenamefont {Khudik}, \citenamefont {Kalmykov},\ and\ \citenamefont
  {Shvets}}]{Yi2011}%
  \BibitemOpen
  \bibfield  {author} {\bibinfo {author} {\bibfnamefont {S.~A.}\ \bibnamefont
  {Yi}}, \bibinfo {author} {\bibfnamefont {V.}~\bibnamefont {Khudik}}, \bibinfo
  {author} {\bibfnamefont {S.~Y.}\ \bibnamefont {Kalmykov}}, \ and\ \bibinfo
  {author} {\bibfnamefont {G.}~\bibnamefont {Shvets}},\ }\bibfield  {title}
  {\enquote {\bibinfo {title} {{Hamiltonian analysis of electron self-injection
  and acceleration into an evolving plasma bubble}},}\ }\href {\doibase
  10.1088/0741-3335/53/1/014012} {\bibfield  {journal} {\bibinfo  {journal}
  {Plasma Physics and Controlled Fusion}\ }\textbf {\bibinfo {volume} {53}},\
  \bibinfo {pages} {014012} (\bibinfo {year} {2011})}\BibitemShut {NoStop}%
\bibitem [{\citenamefont {Li}\ \emph {et~al.}(2015)\citenamefont {Li},
  \citenamefont {Yu}, \citenamefont {Gu}, \citenamefont {Huang}, \citenamefont
  {Kong},\ and\ \citenamefont {Kawata}}]{Li2015}%
  \BibitemOpen
  \bibfield  {author} {\bibinfo {author} {\bibfnamefont {X.~F.}\ \bibnamefont
  {Li}}, \bibinfo {author} {\bibfnamefont {Q.}~\bibnamefont {Yu}}, \bibinfo
  {author} {\bibfnamefont {Y.~J.}\ \bibnamefont {Gu}}, \bibinfo {author}
  {\bibfnamefont {S.}~\bibnamefont {Huang}}, \bibinfo {author} {\bibfnamefont
  {Q.}~\bibnamefont {Kong}}, \ and\ \bibinfo {author} {\bibfnamefont
  {S.}~\bibnamefont {Kawata}},\ }\bibfield  {title} {\enquote {\bibinfo {title}
  {{Bubble shape and electromagnetic field in the nonlinear regime for laser
  wakefield acceleration}},}\ }\href {\doibase 10.1063/1.4928908} {\bibfield
  {journal} {\bibinfo  {journal} {Physics of Plasmas}\ }\textbf {\bibinfo
  {volume} {22}} (\bibinfo {year} {2015}),\ 10.1063/1.4928908}\BibitemShut
  {NoStop}%
\bibitem [{\citenamefont {Horn{\'{y}}}(2018)}]{horny2018}%
  \BibitemOpen
  \bibfield  {author} {\bibinfo {author} {\bibfnamefont {V.}~\bibnamefont
  {Horn{\'{y}}}},\ }\emph {\bibinfo {title} {Generation of X-Rays by Laser
  Accelerated Electron Beam}},\ \href@noop {} {Ph.D. thesis},\ \bibinfo
  {address} {Czech Technical University in Prague} (\bibinfo {year}
  {2018})\BibitemShut {NoStop}%
\bibitem [{\citenamefont {Davoine}\ \emph {et~al.}(2008)\citenamefont
  {Davoine}, \citenamefont {Lefebvre}, \citenamefont {Faure}, \citenamefont
  {Rechatin}, \citenamefont {Lifschitz},\ and\ \citenamefont
  {Malka}}]{davoine_pop2008}%
  \BibitemOpen
  \bibfield  {author} {\bibinfo {author} {\bibfnamefont {X.}~\bibnamefont
  {Davoine}}, \bibinfo {author} {\bibfnamefont {E.}~\bibnamefont {Lefebvre}},
  \bibinfo {author} {\bibfnamefont {J.}~\bibnamefont {Faure}}, \bibinfo
  {author} {\bibfnamefont {C.}~\bibnamefont {Rechatin}}, \bibinfo {author}
  {\bibfnamefont {A.}~\bibnamefont {Lifschitz}}, \ and\ \bibinfo {author}
  {\bibfnamefont {V.}~\bibnamefont {Malka}},\ }\bibfield  {title} {\enquote
  {\bibinfo {title} {{Simulation of quasimonoenergetic electron beams produced
  by colliding pulse wakefield acceleration}},}\ }\href {\doibase
  10.1063/1.3008051} {\bibfield  {journal} {\bibinfo  {journal} {Physics of
  Plasmas}\ }\textbf {\bibinfo {volume} {15}},\ \bibinfo {pages} {113102}
  (\bibinfo {year} {2008})}\BibitemShut {NoStop}%
\end{thebibliography}%
\end{document}